\documentclass[sn-mathphys-num]{sn-jnl}
\catcode`\|=12\relax
\let\orcidlogo\relax
\usepackage{orcidlink}
\usepackage{graphicx}
\usepackage{tikz} 
\usepackage[]{framed}
\usepackage{amsmath}
\usepackage{adjustbox}
\usepackage{xspace}
\usepackage{soul}
\usepackage{placeins}  
\usepackage{longtable}
\usepackage{float}
\usepackage[inline]{enumitem}
\usepackage{booktabs}
\usepackage{array, multirow, multicol}
\usepackage[normalem]{ulem}
\usepackage{geometry}
\usepackage{tablefootnote}
\usepackage{threeparttable}
\usepackage{fancyvrb}
\usepackage{xcolor}
\definecolor{LightGray}{gray}{0.9}
\usepackage{tcolorbox}
\tcbuselibrary{listings,breakable}

\newcommand\mmarco{\textsc{mMARCO}\xspace}
\newcommand\msmarco{\textsc{MS MARCO}\xspace}
\newcommand\quati{\textsc{Quati}\xspace}
\newcommand\chave{\textsc{CHAVE}\xspace}
\newcommand\juristcu{\textsc{JurisTCU}\xspace}
\newcommand\thisdataset{\textsc{NormasTCU}\xspace}
\newcommand\ulyssesrfcorpus{\textsc{Ulysses-RFCorpus}\xspace}

\begin{document}
\let\WriteBookmarks\relax

\title[\textsc{NormasTCU} --- A Brazilian Portuguese IR Dataset and an Evaluation of LLM-as-a-Judge for Relevance Assessment]{\textsc{NormasTCU} --- A Brazilian Portuguese IR Dataset and an Evaluation of LLM-as-a-Judge for Relevance Assessment}

\author*[1]{\fnm{Leandro} \sur{Carísio Fernandes}\orcidlink{0000-0002-4114-2334}}\email{carisio@gmail.com}
\author[2]{\fnm{Marcus} \sur{Vinícius Borela de Castro}\orcidlink{0009-0000-8633-1385}}\email{marcusborela@yahoo.com.br}
\author[2]{\fnm{Leandro} \sur{dos Santos Ribeiro}\orcidlink{0009-0008-3715-9927}}\email{leandro.santos.r@gmail.com}
\author[2]{\fnm{Leonardo} \sur{Augusto da Silva Pacheco}\orcidlink{0000-0001-6506-2366}}\email{leonardo3108@gmail.com}
\author*[2]{\fnm{Edans} \sur{Flávius de Oliveira Sandes}\orcidlink{0000-0001-9502-7799}}\email{edansfs@tcu.gov.br}

\affil[1]{\orgname{Câmara dos Deputados}, \orgaddress{\city{Brasília}, \country{Brazil}}}
\affil[2]{\orgname{Tribunal de Contas da União (TCU)}, \orgaddress{\city{Brasília}, \country{Brazil}}}

\abstract{
Portuguese Information Retrieval (IR) lacks public datasets, and relevance assessment for specialized collections remains costly. While Large Language Models (LLMs) increasingly support relevance assessment, their reliability in non-English specialized domains remains unclear. We introduce \textsc{NormasTCU}\hyperlink{fn:dataset}{\textsuperscript{1}}, a Brazilian Portuguese IR dataset with 14,469 legal documents, 46 queries, and 3,048 human judgments over 812 query-document pairs. Using \thisdataset, we evaluated LLM-as-a-judge for relevance assessment by prompting three models with two prompt techniques to grade these pairs. We then compared the rankings of 15 IR systems derived from LLM-generated and human reference qrels. LLMs consistently showed a positive scoring bias (mean absolute error: 0.46--0.66 on a 0-2 scale). Furthermore, pair-level agreement with human judgments achieved only fair to moderate levels, with Cohen's $\kappa$ ranging from 0.32 to 0.53. Despite this bias, LLM-generated judgments often yielded highly similar system rankings for nDCG@10 and MRR (observed Kendall's $\tau \geq$ 0.90, although the bootstrap confidence intervals did not always remain above this threshold), but were less reliable for P@10 and R@10. Notably, LLM-based rankings were sometimes more strongly correlated with the reference ranking than individual human annotations were. As a practical implication, our results suggest that LLMs could effectively support scalable relevance assessment in specialized Portuguese corpora when evaluated using nDCG or MRR (rank-aware metrics), but they should be avoided when relying on precision or recall.}



\maketitle


\stepcounter{footnote}%
\footnotetext{\hypertarget{fn:dataset}{}\url{https://huggingface.co/datasets/LeandroRibeiro/NormasTCU}}

\section{Introduction}

Information Retrieval (IR) systems are usually evaluated using offline test collections \cite{sanderson2010test}, which, under the Cranfield paradigm, consist of a document corpus, a set of queries (topics), and relevance judgments (qrels) for query-document pairs \cite{Cranfield_tests_on_index_language_devices,voorhees2001philosophy}. Currently, most publicly available IR test collections are in English, which limits the evaluation of retrieval systems in other languages. In particular, Portuguese is highly relevant, as it is one of the world's most widely spoken languages\footnote{\url{https://en.wikipedia.org/wiki/List_of_languages_by_total_number_of_speakers}}.

Retrieval performance also depends on factors beyond language, such as the nature of the document collection. Domain-specific corpora introduce challenges that are not present in general-purpose datasets. Among these, the legal domain stands out as a key scenario. Legal professionals rely on IR applications for case law retrieval and document analysis; moreover, legal search is essential for transparency, decision-making, and access to justice in public institutions. Adapting IR systems to this domain is non-trivial, as, unlike general web corpora, official legal documents tend to be highly structured and rely on specialized vocabulary and strict terminology \cite{lir_state_of_the_art, NGUYEN2025103949, YANG2026104816}. Thus, properly evaluating these systems requires domain-specific test collections, which remain scarce in Portuguese.

Building a Cranfield-style collection is costly. Ideally, relevance would be assessed manually for the entire corpus by trained human assessors following detailed guidelines. In practice, exhaustive judging is rarely feasible today due to the volume of modern corpora. Therefore, most collections rely on assessing only a subset of documents. Even then, the process remains time-consuming \cite{voorhees2001philosophy, sanderson2010test}.

To mitigate these costs, alternatives to manual assessment are currently being investigated in the IR community. In this context, large language models (LLMs) have been used to evaluate complex tasks, a paradigm known as LLM-as-a-judge \cite{gu2025surveyllmasajudge}. One such task is document relevance annotation, which offers a scalable solution by enabling the annotation of large document collections at substantially lower cost than manual assessment. Prior studies have shown that LLM-generated judgments introduce bias at the query-document pair level, usually assigning higher relevance scores than human annotators \cite{alaofi2026llmforrelevancelabelling, rahmani2025judgingjudges, lim_auto_rele_ass_llm, yu2026llmjudgesinflatescores}. Nevertheless, when LLM-generated relevance judgments are used to evaluate IR systems, the resulting system rankings are often highly correlated with those obtained from human judgments \cite{rahmani2025judgingjudges, upadhyay2024largescale, upadhyay2024umbrela}. That is, although LLMs may diverge from humans in absolute relevance scores, they frequently preserve the relative ordering of IR systems.

However, most publicly available IR test collections, and consequently most empirical evidence on the quality of LLM-generated annotations, are centered on English and on everyday language data. Since these models are predominantly trained on English-language data, their behavior in relevance assessment may vary across languages and domains \cite{WANG2026104616}, and the extent to which findings generalize beyond English remains an open question. Thus, evaluating the use of LLM annotators in other languages and domains is necessary to determine whether their performance remains consistent.

To address these gaps, this paper presents a two-fold contribution. First, we introduce \thisdataset, a new Brazilian Portuguese IR test collection with 14,469 legal documents, 46 queries, and graded relevance judgments produced by four human annotators. The corpus consists of normative acts issued by the Brazilian Federal Court of Accounts, while the queries reflect real user search terms. We publicly released the corpus, queries, and relevance judgments (provided as aggregated qrels and anonymized individual annotations) to help address the scarcity of Portuguese IR resources in the legal domain.

Second, considering that \thisdataset was built using human judgments, we conduct an experiment to assess how closely LLM-generated relevance annotations align with human judgments in Portuguese for ranking IR systems. We analyzed agreement between human and LLM assessments at two levels: (i) the query-document pair level, evaluating agreement and bias; and (ii) the system level, examining whether LLM-generated qrels preserve the ranking of IR systems obtained with aggregated human judgments. Unlike recent studies that typically rely on nDCG \cite[Normalized Discounted Cumulative Gain, e.g.,][]{rahmani2025judgingjudges, bencke2024trust, upadhyay2024largescale}, we evaluated multiple metrics to investigate whether ranking consistency generalizes across them. We also compared rankings derived from LLM-generated qrels with those based on individual human annotators.

The experiment on the \thisdataset dataset, using the prompts and models evaluated in this study, showed that LLMs tend to assign higher relevance scores than humans, i.e., there is a positive bias at the query-document pair level. Nevertheless, system rankings derived from LLM-generated qrels remained highly correlated with those obtained using aggregated human judgments when evaluated with nDCG@10 and MRR (Mean Reciprocal Rank) metrics, but not with P@10 (Precision) or R@10 (Recall). Notably, in many cases during the experiment, due to variability among human annotators, the correlation between LLM-based rankings and the aggregated human ranking was higher than the correlation between a single human annotator's ranking and the aggregated human ranking.

\subsection{Research objectives (RO)}
\label{sec:research_objectives}

We now formally define the research objectives that guided this study.

\begin{itemize}
    \item \textbf{RO1:} To develop and publicly release \thisdataset, a Brazilian Portuguese legal IR test collection that includes graded relevance judgments over query-document pairs provided independently by four domain experts.

    \item \textbf{RO2:} To re-annotate the query-document pairs of \thisdataset using different LLM/prompt combinations and to compare the quality of these automated assessments against human judgments, identifying potential biases in LLM-generated annotations in a specialized, non-English domain.

    \item \textbf{RO3:} To determine the downstream impact of LLM-generated judgments on IR system evaluation by assessing whether they preserve the relative ranking of retrieval systems across several evaluation metrics, even when pair-level annotations diverge.
\end{itemize}

\subsection{Contributions}

Considering the current state of research in this field, the main contributions of this work are as follows:

\begin{itemize}
\item We introduce a new Brazilian Portuguese IR test collection (\thisdataset), which includes a corpus of legal documents and qrels generated from four independent human annotators.
\item We also advance the study of LLM-as-a-judge in Portuguese by releasing anonymized individual human annotations and conducting an experiment that compares rankings derived from the reference qrels with those from LLM-generated qrels.
\end{itemize}

\section{Related work}

Portuguese IR resources are still limited compared to English. Early efforts include the \chave collection, created in the context of CLEF (Conference and Labs of the Evaluation Forum) in 2004. It contains about 16,000 full newspaper articles from 1994 to 1995, 100 queries, and human-generated relevance judgments \cite{chave2004}.

More recently, datasets have been translated into Portuguese. For example, \mmarco was created by translating English passages (short snippets of text) and queries from \msmarco into thirteen languages, including Portuguese \cite{bonifacio2021mmarco}. It provides large-scale training and evaluation data for passage retrieval, with a corpus of about 8.8 million passages.

Datasets natively created in Brazilian Portuguese remain scarce; however, a few datasets have recently been proposed to support research in Portuguese IR, namely \quati, \ulyssesrfcorpus, and \juristcu.

\quati has a corpus of approximately 10 million passages derived from websites in the ClueWeb22 Category B dataset \cite{bueno2024quati}. It provides 200 queries annotated with an LLM-based annotation pipeline.

In the legal domain, dedicated resources have also been developed. The \ulyssesrfcorpus is a relevance feedback corpus designed for IR of legislative documents \cite{camara_dep_ir} and annotated by domain experts. The corpus contains 105,681 documents and 693 queries. Its queries are formulated as requests for legislative proposals (e.g., bills), rather than keyword-based or question-based search queries.

Also within the legal domain, \juristcu is a legal IR test collection built from jurisprudential documents \cite{fernandes2026juristcu}. It includes 16,045 full legal decisions (rulings) and 150 queries, formulated as keyword-based and question-based search queries. Relevance judgments were obtained through a hybrid method in which LLM-generated scores were reviewed by a human annotator.

Beyond the characteristics of the datasets themselves, another important aspect concerns how relevance judgments are obtained. While collections like \chave and \ulyssesrfcorpus rely entirely on human experts, LLM-based approaches were used to generate the qrels for \quati and \juristcu. In this context, the authors of \quati sampled a subset of the automatically generated labels, performed manual assessments, and reported a Cohen's $\kappa$ of 0.31 on a four-grade scale \cite{bueno2024quati}. In the case of \juristcu, the review of LLM-generated scores by a domain expert was fully consistent with the automatic judgments \cite{fernandes2026juristcu}. The authors argued that expert validation of pre-generated labels can introduce confirmation bias, and unless the output is highly inconsistent (e.g., a clearly irrelevant document labeled as highly relevant), annotators tend to agree with the judgments they are reviewing.

Recent studies have evaluated the use of LLMs for relevance assessment and suggest that their judgments exhibit systematic biases at the query-document pair level, often assigning higher relevance scores than human annotators \cite{alaofi2026llmforrelevancelabelling, rahmani2025judgingjudges, lim_auto_rele_ass_llm, yu2026llmjudgesinflatescores}. However, since relevance judgments are ultimately used to rank and compare retrieval systems, it is also necessary to examine the stability of system rankings produced by different sets of qrels. When examined at the system level, evaluations based on LLM-generated qrels frequently produce rankings that are highly correlated with those derived from human judgments \cite{rahmani2025judgingjudges, upadhyay2024largescale, upadhyay2024umbrela}. In other words, LLM-based assessment tends to preserve the relative ordering of retrieval systems even when absolute relevance scores differ.

Most evidence in this field so far comes from English benchmarks. For this reason, evaluations in other languages are also necessary to verify whether pair-level bias persists and whether system rankings remain stable beyond English. We found one study that evaluated the use of LLM annotators at the system level using a collection created natively in Portuguese (\chave), which reported similar conclusions; however, it tested only two models across four ranking functions \cite{bencke2024trust}.

\subsection{Distinction from existing work}

Our contributions differ from and complement prior work in several aspects. Regarding the dataset itself, in addition to being based on a different corpus than \ulyssesrfcorpus and \juristcu, relevance judgments in \thisdataset were independently assigned by four domain experts to the same query-document pairs. In contrast, in \ulyssesrfcorpus, each query-document pair is assessed by a single annotator, whereas \juristcu relies on automatically generated relevance labels. Moreover, unlike \ulyssesrfcorpus, whose queries are formulated as requests for legislative documents, the queries in \thisdataset represent document search queries. Finally, by releasing the individual assessments, \thisdataset enables the study of inter-annotator agreement and judgment variability. Table~\ref{tab:dataset_comparison} summarizes the main differences among the datasets.

In terms of the LLM-as-a-judge evaluation, evidence for Portuguese IR remains limited. To the best of our knowledge, the only previous system-level evaluation in Portuguese considered two LLMs and four retrieval functions \cite{bencke2024trust}. We extend this analysis by evaluating three LLMs, two prompting strategies, and rankings generated from fifteen retrieval systems. This provides a broader assessment of the ability of LLM-generated qrels to preserve IR system rankings for Portuguese collections.

\begin{table}[!htb]
\renewcommand{\arraystretch}{1.2}
\centering
\caption{Comparison between other Portuguese IR datasets and \thisdataset.}
\label{tab:dataset_comparison}
\begin{tabular}{p{2cm}p{4cm}p{1cm}p{1cm}p{5cm}}
\toprule
\textbf{Dataset} & \textbf{Domain} & \textbf{Corpus} & \textbf{Queries} & \textbf{Qrels (annotated by)} \\
\midrule
\chave & News & $\sim$16k & 100 & Human \\
\mmarco & Web passages & $\sim$8.8M & 6.980 & Human \\
\quati & Web passages & $\sim$10M & 200 & LLM \\
\ulyssesrfcorpus & Legal (legislative documents) & 105,681 & 693 & Domain experts (one for each query-document pair) \\
\juristcu & Legal (jurisprudence) & 16,045 & 150 & LLM and reviewed by one domain expert \\
\thisdataset & Legal (normative acts) & 14,469 & 46 & Four domain experts for each query-document pair \\
\bottomrule
\end{tabular}
\end{table}

\section{\textsc{NormasTCU} -- Dataset construction}

Addressing the first research objective (\textbf{RO1}), this section details the construction of \thisdataset, a new test collection for legal information retrieval in Brazilian Portuguese, with a corpus of 14,469 normative documents, 46 user queries, and relevance judgments graded on a three-level scale by four independent domain experts.

The Brazilian Federal Court of Accounts (Tribunal de Contas da União -- TCU) maintains collections of institutional documents. In this paper, we focus on one of these resources: a collection of normative documents that produce either internal effects (e.g., rules governing staff and internal procedures) or external effects (e.g., provisions regulating the Court's interactions with other public bodies). These documents are part of a broader institutional database that includes both public and private records; here, we consider only the public subset. The original annotations were conducted for institutional use to improve internal information retrieval systems; however, for research and reproducibility, we filtered out all private documents and their associated judgments to release the \thisdataset collection.

\subsection{Document corpus}

TCU provides its public normative corpus in CSV format\footnote{\url{https://pesquisa.apps.tcu.gov.br/dados-abertos}}, updated weekly. We used the July 2024 snapshot as the corpus of \thisdataset, totaling 14,469 public documents, each written in Brazilian Portuguese and corresponding to a single normative act.

Documents are structured into multiple fields, shown in Table~\ref{tab:dataset_docs}; among them, the most relevant textual fields are TEXTONORMA, which is the full text of the normative act, and ASSUNTO, which is a brief summary of its content (hereafter referred to as TEXT and SUMMARY, respectively). Table~\ref{tab:text_cdf} reports the percentiles of the number of characters in the TEXT field, for both the original content and the HTML-stripped version. Overall, the texts are short: after stripping HTML, 75\% have fewer than 3,475 characters. However, the corpus also contains long documents: more than 10\% exceed 11,067 characters.

\begin{table}[htbp]
\caption{Document structure.}
\renewcommand{\arraystretch}{1.2}
\centering
{%
\begin{tabular}{p{4cm}p{5.3cm}p{1.7cm}p{2.4cm}}
\toprule
\textbf{Field} & \textbf{Description} & \textbf{Type} & \textbf{Missing values} \\ \hline
KEY & ID & Text & -- \\
UNIDADEBASICAAUTORA & Authoring organizational unit & Text & -- \\
ORIGEM & Source of the normative act & Text & -- \\
NUMNORMA & Document number & Text & -- \\
ANONORMA & Document year & Text & -- \\
TIPONORMA & Document type & Text & -- \\
TITULO & Document title & Text & -- \\
ASSUNTO & Subject / summary & Text & 7,191 (49.70\%) \\
TEXTONORMA & Full text & HTML Text & 5 (0.03\%) \\
DATAINICIOVIGENCIA & Effective start date & \verb|DD\MM\YYYY| & -- \\
DATAFIMVIGENCIA & Effective end date & \verb|DD\MM\YYYY| & 13,629 (94.19\%) \\
SITUACAO & Expressly revoked or in force & Text & -- \\
TEXTOANEXO & Annex of the document & HTML Text & 13,237 (91.49\%) \\
TEMA & Subject: external control/management & Text & 3,686 (25.48\%) \\
TAGSVCE & Tags & Text & 3,849 (26.60\%) \\
NORMARELACIONADA & Related documents of this dataset & Text & 9,918 (68.55\%) \\
\bottomrule
\end{tabular}%
}
\vspace{0.05cm}
\label{tab:dataset_docs}
\end{table}

\begin{table}[htbp]
\centering
\caption{Percentiles of the number of characters in the TEXT field, considering the original content and the HTML-stripped version.}
\label{tab:text_cdf}
\renewcommand{\arraystretch}{1.2}
\begin{tabular}{lrr}
\toprule
\textbf{Percentile} & \textbf{TEXT} & \textbf{TEXT (HTML-stripped)} \\
\midrule
5  & 991     & 312    \\
25 & 1,500   & 451    \\
50 & 2,504   & 923    \\
75 & 8,525   & 3,475  \\
90 & 28,787	 & 11,067 \\
95 & 57,917  & 19,508 \\
99 & 391,809 & 95,384 \\
\bottomrule
\end{tabular}
\end{table}

\subsection{Queries}

To build the set of queries, we selected terms from the most frequent user queries in the TCU search system over a 12-month period (June 2023 to June 2024). Since the system is BM25-like, queries are usually expressed as short keyword sequences. Prior to the relevance assessment stage, we manually rewrote some of them to make them semantically richer, replacing keyword-based formulations with more descriptive phrasal expressions. In total, we released 46 queries, shown in Table~\ref{tab:queries} in \ref{sec:anexo_queries}, along with their English translations, covering topics ranging from administration and governance to staff regulations and human resources, as well as audit and external control. All subsequent stages of the study (pooling, human assessment, and LLM assessment) used the rewritten queries exactly as released in \thisdataset.

\subsection{Qrels (relevance judgments)}
\label{sec:dataset_qrels}

\begin{figure}[!htbp]
    \centering
    \includegraphics[width=1\linewidth]{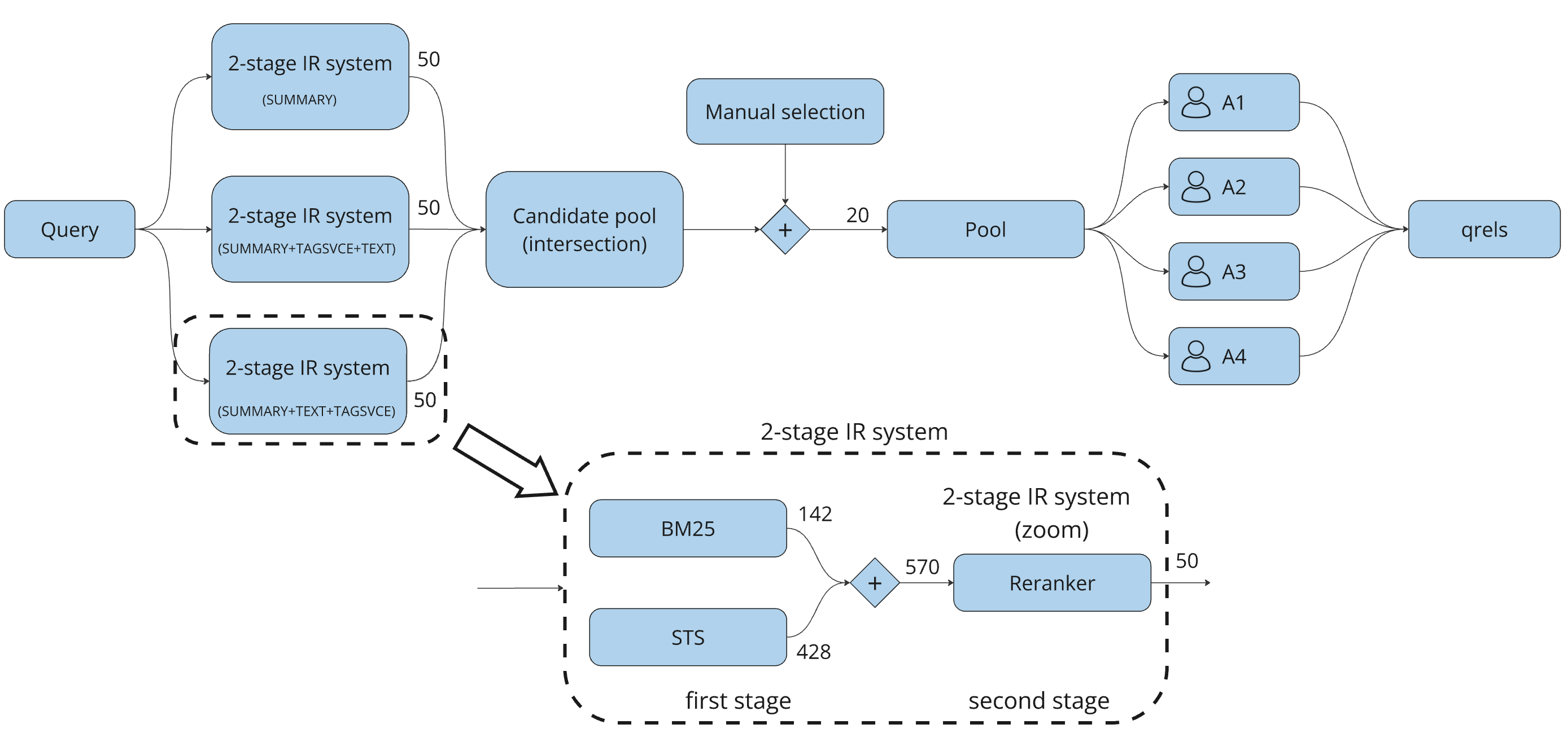}
    \caption{Pipeline to generate the qrels.}
\label{fig:pipeline_qrels}
\end{figure}

We used the pipeline shown in Figure~\ref{fig:pipeline_qrels} to generate the relevance judgments of query-document pairs. For each query in Table~\ref{tab:queries}, we applied a pooling strategy to select 20 candidate documents for annotation \cite{manning2008introduction}.

Each query was submitted to three two-stage IR systems that shared the same configuration but had different corpus representations (different indexes). In the first stage, a hybrid method retrieved 570 candidates for reranking by combining BM25 with semantic textual similarity (STS) using sentence embeddings from the \texttt{rufimelo/Legal-BERTimbau-sts-large-ma-v3} model\footnote{\url{https://huggingface.co/rufimelo/Legal-BERTimbau-sts-large-ma-v3}}. We applied a 25\%:75\% split (BM25/STS), yielding 142 documents retrieved via BM25 and 428 via STS. These hyperparameters (the 570 candidates and the 25\%:75\% distribution) were established through prior empirical evaluations on internal TCU datasets.

The three indexes differed in the textual fields used: (1) \texttt{SUMMARY} alone; (2) \texttt{SUMMARY} + \texttt{TAGSVCE} + \texttt{TEXT}; and (3) \texttt{SUMMARY} + \texttt{TEXT} + \texttt{TAGSVCE}. The latter two indexes contain the same fields, but the tags are arranged in a different order. Although this ordering does not affect BM25, it impacts the embeddings used for semantic retrieval due to word order and truncation effects when the input exceeds the model's context window.

In the second stage, the 570 retrieved candidates were reranked using the \texttt{unicamp-dl/mt5-3B-mmarco-en-pt}\footnote{\url{https://huggingface.co/unicamp-dl/mt5-3B-mmarco-en-pt}} neural reranking model, a Portuguese cross-encoder trained on the mMARCO dataset.

For each query, the three two-stage pipelines produced ranked lists of 50 documents. We first selected the documents that appeared in all three lists (i.e., the intersection) and then completed the pool of 20 documents per query by manually choosing additional candidates. This manual selection was performed by the authors, either from the non-overlapping documents returned by the retrieval systems or, when necessary, through manual searches in the TCU retrieval system. These 20 documents per query formed the final pool sent for human assessment.

\begin{figure}[!htbp]
    \centering
    \includegraphics[width=0.9\linewidth]{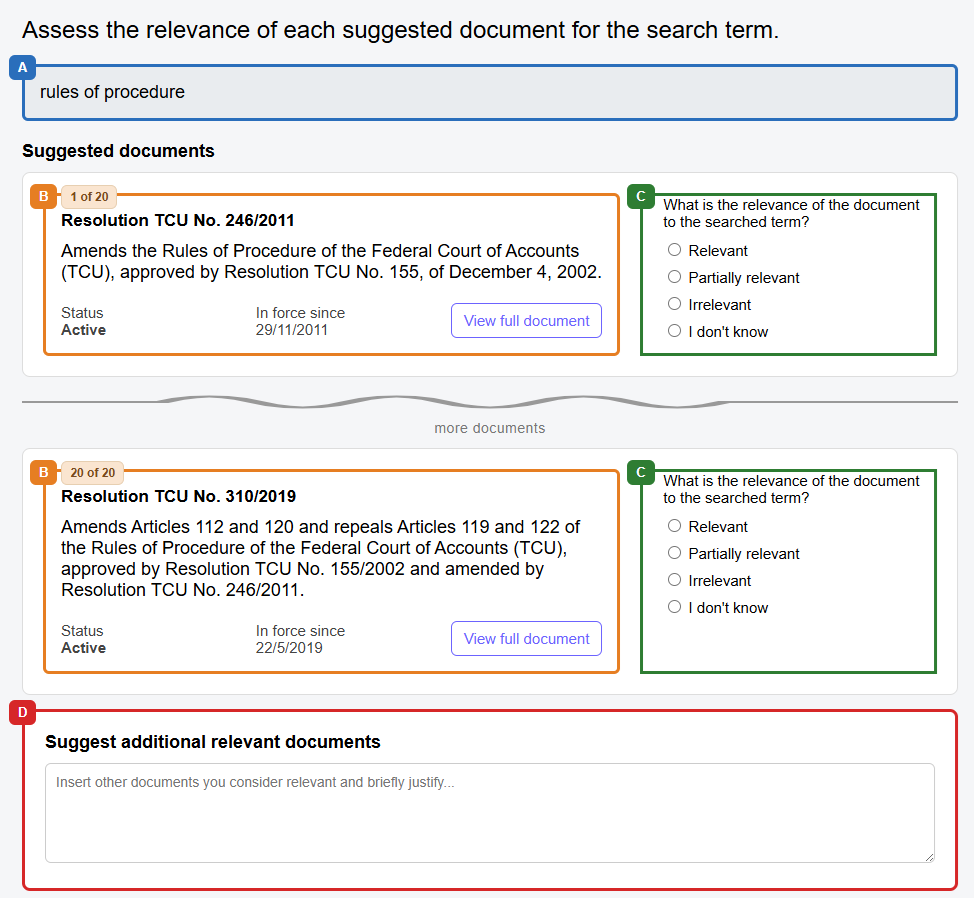}
    \caption{Screenshot of the annotation system used for relevance assessment (adapted from the original tool for presentation in this article). The interface shows (A) the query; (B) the document title and summary, with the option to open the full document; (C) relevance grading options for the query-document pair; and (D) an option to provide additional relevant documents.}
\label{fig:screenshot}
\end{figure}

We selected four domain experts to independently assess the same pool of 20 query-document pairs using a three-level relevance scale (\texttt{irrelevant}, \texttt{partially relevant}, and \texttt{relevant}). Annotators could also indicate that they were unable to assess a query-document pair (marked as \texttt{cannot judge}). The evaluation was conducted using a custom application developed for this task, shown in Figure~\ref{fig:screenshot}, and took approximately one month to complete.

All annotators were TCU auditors from different organizational units and were involved in the curation of the normative documents used in this study. Thus, they were familiar with the collection and its contents. Annotation guidelines were provided verbally during a meeting attended by all annotators. The relevance labels were defined as follows:

\begin{itemize}
    \item \texttt{relevant}: the document directly answers the query and contains the core information requested, providing sufficient information for the user to satisfy their search objective.

    \item \texttt{partially relevant}: the document mentions the topic of the query or addresses a related subject, but does so only partially or insufficiently to fully satisfy the information need.

    \item \texttt{irrelevant}: the document does not address the subject of the query.
\end{itemize}

The annotators were informed that the purpose of the annotation effort was to build a test collection to evaluate the organization's internal search system and support future improvements. They shared the understanding that an ideal search system should rank all relevant documents before partially relevant ones. They were also informed that the interpretation of each query's intent should be based solely on their own judgment.

Test collections constructed through pooling may leave some relevant documents unjudged and introduce bias in favor of the systems used during pooling. However, previous studies have shown that pools built with adequate depth and diversity can still provide reliable comparative evaluations of retrieval systems \cite{zobel1998reliable, voorhees2001philosophy}. In our case, the annotators were domain experts who curated the document collection and were familiar with its contents. They were also allowed to add documents they considered relevant that were not included in the initial pool (box D of Figure \ref{fig:screenshot}). These design choices helped reduce the impact of missing relevant documents.

In total, only seven new query-document pairs were added. These pairs were considered \texttt{relevant} or \texttt{partially relevant} by the annotator who suggested them, based on their justification, and were not assessed by the other annotators. The small number of additional pairs suggests that the pooling strategy successfully selected the most potentially relevant documents.

Using these assessments, we built the qrels for \thisdataset by averaging the annotators' scores for each query-document pair. The relevance labels were mapped to scores of 0-\texttt{irrelevant}, 1-\texttt{partially relevant}, and 2-\texttt{relevant}. Pairs marked as \texttt{cannot judge} were excluded only for the respective annotator during aggregation. As averaging may result in fractional values and standard evaluation tools generally assume integer labels (e.g., Pyserini \cite{Lin_etal_SIGIR2021_Pyserini} and pytrec\_eval \cite{Van_Gysel_2018}), the final scores were rounded using Banker's rounding.

Because we conducted an experiment comparing different qrels, the qrels for \thisdataset are hereafter referred to as the \textit{reference qrels}. In total, this collection contains 812 unique judged query-document pairs across 46 queries, corresponding to 3,048 individual relevance judgments, which are also released in anonymized form.

It is important to note that the annotators judged the queries without being given an explicit statement of the underlying search intent. One advantage of this design is that it more closely mirrors real-world search scenarios, in which users typically submit short queries and the IR system ranks documents without knowing the actual user intent. A drawback is that different annotators may interpret the same query differently, which can increase disagreement among them and potentially penalize documents that would be considered \texttt{relevant} under alternative interpretations.

\subsection{Quality of human annotations}

Table~\ref{tab:label_distribution_annotators} reports the distribution of the 3,048 individual human judgments across relevance labels for the query-document pairs. Although the dataset contains 812 unique query-document pairs, no annotator evaluated all of them. Of the 3,048 pairs, 63\% were judged \texttt{irrelevant}. However, relevance label distributions differ across annotators.

\begin{table}[!htb]
\renewcommand{\arraystretch}{1.2}
\centering
\caption{Label distribution per annotator (counts).}
\label{tab:label_distribution_annotators}
\begin{tabular}{lrrrr||rr}
\toprule
\textbf{Annotator} & \textbf{Irrelevant} & \textbf{Partially relevant} & \textbf{Relevant} & \textbf{\shortstack{Judged\\ pairs}} & \textbf{\shortstack{Unjudged\\pairs}} & \textbf{\shortstack{Added\\pairs}} \\
\midrule
$A_1$   & 536 (67\%) & 66 (8\%) & 203 (25\%) & 805 & 7 & 1 \\
$A_2$   & 477 (67\%) & 80 (11\%) & 156 (22\%) & 713 & 99 & 2 \\
$A_3$   & 461 (57\%) & 137 (17\%) & 207 (26\%) & 805 & 7 & -- \\
$A_4$   & 451 (62\%) & 166 (23\%) & 108 (15\%) & 725 & 87 & 4 \\
\midrule
\textbf{Total} & 1925 (63\%) & 449 (15\%) & 674 (22\%) & 3048 & 200 & 7 \\
\bottomrule
\end{tabular}
\end{table}

Table~\ref{tab:pair_level_agreement} shows the percentage of agreement among the four annotators at the query-document pair level. For this analysis, we considered only the available judgments for each pair. A \textit{consensus} is reached when all annotators who assessed the pair assigned the same relevance score. More than half of the pairs (54\%) achieved full consensus, and this agreement is mostly driven by the \texttt{irrelevant} label (45\% of all pairs). Beyond consensus, 28\% of the pairs show a disagreement of at most one relevance level among annotators (0-1 or 1-2), which we refer to as \textit{weak disagreement}. Finally, 18\% of the pairs show \textit{strong disagreement} among the available annotators (scores of both 0 and 2 are assigned to the same pair).

\begin{table}[!htb]
\renewcommand{\arraystretch}{1.2}
\centering
\caption{Pair-level agreement and disagreement across the four human annotators (all judged pairs).}
\label{tab:pair_level_agreement}
\begin{tabular}{lrr}
\toprule
\textbf{Category} & \textbf{Count} & \textbf{Percent} \\
\midrule
\textbf{Consensus (total)} & \textbf{437} & \textbf{54\%} \\
\quad Score 0 & 360 & 45\% \\
\quad Score 1 &  11 &  1\% \\
\quad Score 2 &  66 &  8\% \\
\midrule
\textbf{Weak disagreement (total)} & \textbf{225} & \textbf{28\%} \\
\quad Scores 0-1 & 135 & 17\% \\
\quad Scores 1-2 &  90 & 11\% \\
\midrule
\textbf{Strong disagreement (0-2)} & \textbf{150} & \textbf{18\%} \\
\bottomrule
\end{tabular}
\end{table}

A certain degree of disagreement is a well-documented phenomenon in IR test collections, as relevance is inherently subjective \cite{voorhees2000variations}. To illustrate how this subjectivity translates into annotation disagreement, consider the first query in Table~\ref{tab:queries} (\textit{regimento interno}, i.e., Rules of Procedure). TCU has two main documents related to this query: one corresponding to the Court's Rules of Procedure and another to its educational institute, an internal unit responsible for training and professional development; both have been amended by several normative acts. Consequently, assessors may diverge when judging documents for this query: one may judge any related acts as \texttt{relevant}; another may restrict the \texttt{relevant} label strictly to the Rules of Procedure, grading amendments as \texttt{partially relevant}; and another may consider the documents related to the educational institute as \texttt{irrelevant}. This shows that some of the observed variance results from legitimate differences in semantic interpretation rather than from annotation noise or a lack of expertise.

\begin{figure}[!htbp]
    \centering
    \includegraphics[width=0.377\linewidth]{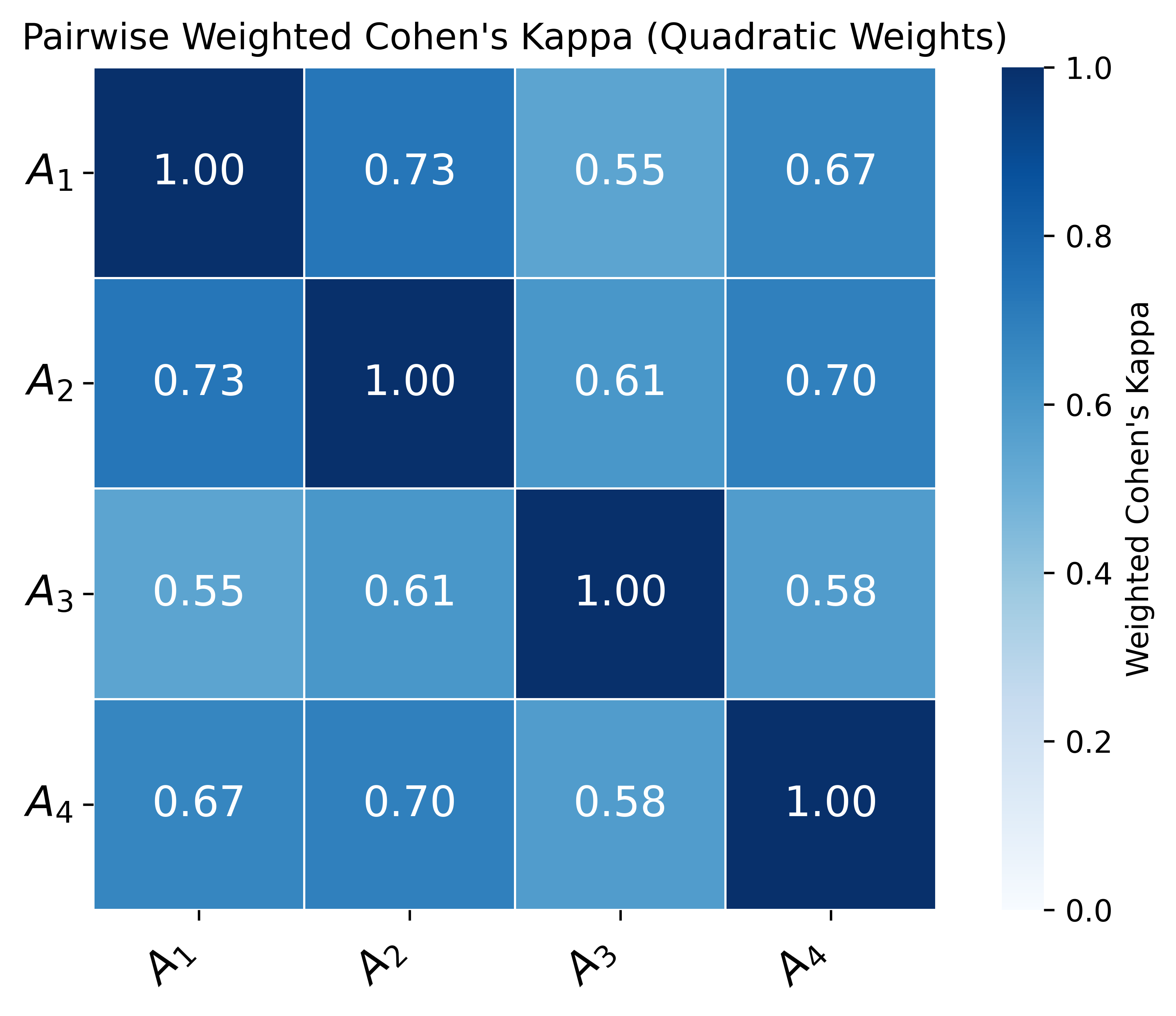}
    \caption{Pairwise weighted Cohen's $\kappa$ (quadratic) between human annotators.}
    \label{fig:pairwise_kappa_human}
\end{figure}

To measure the reliability of the annotations, Figure~\ref{fig:pairwise_kappa_human} shows the pairwise quadratic weighted Cohen's $\kappa$ agreement between annotators on the three-level relevance scale. According to conventional interpretations of this metric (Table \ref{tab:kappa_interpretation}), the observed agreement ranges from moderate ($\kappa \in [0.41, 0.60]$) to substantial ($\kappa \in [0.61, 0.80]$). The highest agreement was observed between $A_1$ and $A_2$ ($\kappa = 0.73$), whereas the lowest agreement involved $A_3$ and the others (e.g., $A_1$ vs. $A_3$, $\kappa = 0.55$).

\begin{table}[!htb]
\renewcommand{\arraystretch}{1.2}
\centering
\caption{Interpretation of Cohen's $\kappa$ values \cite{landis1977kappa, viera2005kappa, mchugh2012kappa}.}
\label{tab:kappa_interpretation}
\begin{tabular}{ll}
\toprule
\textbf{$\kappa$ range} & \textbf{Strength of Agreement} \\
\midrule
$<0$ & Poor \\
$0.00$--$0.20$ & Slight \\
$0.21$--$0.40$ & Fair \\
$0.41$--$0.60$ & Moderate \\
$0.61$--$0.80$ & Substantial \\
$0.81$--$1.00$ & Almost perfect \\
\bottomrule
\end{tabular}
\end{table}

Although agreement among annotators is not perfect, these values are within ranges considered acceptable in IR test collections, where relevance is inherently subjective and disagreements are expected, especially under graded scales \cite{manning2008introduction, artstein-poesio-2008-survey}.

\section{Evaluating LLM-as-a-judge for IR relevance assessment}

This section presents an experiment designed to evaluate the use of LLM annotators for relevance assessment. Using the \thisdataset dataset, we compare LLM-generated judgments with human annotations and analyze whether these judgments yield consistent IR system rankings.

\subsection{Method}
\label{sec:method}

To address the second research objective (\textbf{RO2}), we prompted LLMs to assign graded relevance labels to the same 812 query-document pairs assessed by human annotators, using the same scale (0-\texttt{irrelevant}, 1-\texttt{partially relevant}, 2-\texttt{relevant}). Multiple model-prompt combinations were evaluated to test sensitivity to both model choice and instruction format, as prior work shows that LLM relevance judgments can vary along both dimensions \cite{thomas2024searcherpref}. Each combination produced a complete set of judgments, which served as LLM-generated qrels.

The models evaluated were \texttt{DeepSeek-V3.2}, \texttt{gpt-5-mini-2025-08-07}, and \texttt{sabiazinho-4-2026-01-06} -- three proprietary models from different providers, the latter being a Brazilian Portuguese LLM \cite{laitz2026sabia4technicalreport}. For readability, the remainder of this paper refers to these models as \texttt{deepseek-3.2}, \texttt{gpt-5-mini}, and \texttt{sabiazinho-4}, respectively. We tested two prompting techniques: \texttt{simple}, where the model receives the query and document to directly output a score; and \texttt{rationale}, where it provides a brief justification (rationale) before assigning the final relevance score. The rationale prompt was included to test whether this technique would improve agreement with human annotators, as previous studies have shown that asking models to justify their answers before responding can improve performance across a variety of reasoning tasks \cite{explainyourself, rationale}. Figures~\ref{fig:prompt_simple} and \ref{fig:prompt_rationale} in \ref{sec:prompts} show both prompts, in Portuguese, and Figures~\ref{fig:prompt_simple_en} and \ref{fig:prompt_rationale_en} show their English translations.

Each combination of model and prompt was treated as an independent LLM annotator. For readability, they are referred to using the naming convention \texttt{\{model\}\_\{prompt\}}, where \texttt{\{model\}} is the LLM used and \texttt{\{prompt\}} is the prompt technique. Table \ref{tab:llm_judges} summarizes the LLM annotators.

\begin{table}[htbp]
\renewcommand{\arraystretch}{1.2}
\caption{LLM annotators used in the experiment.}
\label{tab:llm_judges}
\centering
\begin{tabular}{lll}
\toprule
\textbf{LLM annotator} & \textbf{Model} & \textbf{Prompt technique} \\
\midrule
deepseek-3.2\_simple & DeepSeek-V3.2 & simple \\
deepseek-3.2\_rationale & DeepSeek-V3.2 & rationale \\
gpt-5-mini\_simple & gpt-5-mini-2025-08-07 & simple \\
gpt-5-mini\_rationale & gpt-5-mini-2025-08-07 & rationale \\
sabiazinho-4\_simple & sabiazinho-4-2026-01-06 & simple \\
sabiazinho-4\_rationale & sabiazinho-4-2026-01-06 & rationale \\
\bottomrule
\end{tabular}
\end{table}

All LLM annotators assessed document relevance using only the \texttt{TEXT} field after stripping HTML tags. Both the \texttt{simple} and \texttt{rationale} prompts returned responses in JSON format. When a document exceeded a model's context window, its content was truncated to fit the maximum input length supported by that model. The temperature parameter was fixed at 0 for \texttt{deepseek-3.2} and \texttt{sabiazinho-4}; this parameter is not configurable for \texttt{gpt-5-mini}. All remaining model parameters were left at their default values. Annotations were generated using \texttt{deepseek-3.2} and \texttt{sabiazinho-4} on February 13, 2026, and using \texttt{gpt-5-mini} on February 14, 2026.

After generating the scores for each model and prompt, we compared the LLM-assigned labels against the human judgments at the query-document pair level. We began with a descriptive analysis of the label distributions across all annotators (both human and LLM). Next, we evaluated how often the LLM scores fell within the range of human assessments by calculating agreement percentages under conditions of human \textit{consensus} and \textit{weak disagreement}. We further analyzed the average scoring bias of the LLM annotators relative to human judgments. To conclude the pair-level analysis, we reported the pairwise agreement among all annotators using the quadratic weighted Cohen's $\kappa$.

To address the third research objective (\textbf{RO3}), we assessed the downstream impact of different qrels by comparing system rankings obtained with LLM-generated qrels to those based on the reference qrels. For this purpose, we evaluated a set of fifteen IR systems shown in Table~\ref{tab:ir_ranking_functions}. Their performance under the reference qrels is reported in Table~\ref{tab:system_metrics_summary}, providing a baseline for \thisdataset. These systems were not selected to achieve state-of-the-art effectiveness, but rather to generate varying levels of performance through different retrieval approaches and configurations.

\begin{table}[htbp]
\renewcommand{\arraystretch}{1.2}
\caption{Retrieval systems used in the evaluation.}
\label{tab:ir_ranking_functions}
\centering
\begin{tabular}{p{1cm}p{9cm}p{3cm}}
\toprule
\textbf{Name} & \textbf{Description} & \textbf{Indexed fields} \\
\midrule
 $S_1$ & BM25 with $(k_1, b) = (0.9, 0.4)$ & SUMMARY+TEXT \\
 $S_2$ & BM25 with $(k_1, b) = (0.9, 0.4)$ & TEXT \\
 $S_3$ & BM25 with $(k_1, b) = (0.82, 0.68)$ & SUMMARY+TEXT \\
 $S_4$ & BM25 with $(k_1, b) = (0.82, 0.68)$ & TEXT \\
 $S_5$ & BM25 with $(k_1, b) = (1, 0.3)$ & SUMMARY+TEXT \\
 $S_6$ & BM25 with $(k_1, b) = (1, 0.3)$ & TEXT \\
 $S_7$ & BM25 with $(k_1, b) = (1, 0.9)$ & SUMMARY+TEXT \\
 $S_8$ & BM25 with $(k_1, b) = (1, 0.9)$ & TEXT \\
 $S_9$ & Semantic search. Model: \texttt{text-embedding-3-large} & TEXT\\
 $S_{10}$ & Semantic search. Model: \texttt{text-embedding-3-small} & TEXT \\
 $S_{11}$ & Semantic search. Model: \texttt{gemini-embedding-001} & TEXT \\
 $S_{12}$ & Semantic search. Model: \texttt{Qwen/Qwen3-Embedding-4B} & SUMMARY+TEXT \\
 $S_{13}$ & Semantic search. Model: \texttt{Qwen/Qwen3-Embedding-8B} & SUMMARY+TEXT \\
 $S_{14}$ & Top-50 results from $S_8$ reranked with \texttt{text-embedding-3-small} & TEXT \\
 $S_{15}$ & Top-50 results from $S_8$ reranked with \texttt{gemini-embedding-001} & TEXT \\
\bottomrule
\end{tabular}
\end{table}

\begin{table}[!htb]
\renewcommand{\arraystretch}{1.2}
\centering
\caption{Performance metrics for each system computed on the reference qrels. Values in parentheses indicate the system's ranking position according to that specific metric (rank 1 = best).}
\label{tab:system_metrics_summary}
\begin{tabular}{lcccc}
\toprule
\textbf{System} & \textbf{P@10} & \textbf{R@10} & \textbf{nDCG@10} & \textbf{MRR} \\
\midrule
$S_1$ & 0.1870 {\footnotesize (3)} & 0.3822 {\footnotesize (2)} & 0.3734 {\footnotesize (2)} & 0.5385 {\footnotesize (2)} \\
$S_2$ & 0.1717 {\footnotesize (7)} & 0.3435 {\footnotesize (5)} & 0.3444 {\footnotesize (5)} & 0.5251 {\footnotesize (4)} \\
$S_3$ & 0.1870 {\footnotesize (3)} & 0.3542 {\footnotesize (4)} & 0.3496 {\footnotesize (4)} & 0.4832 {\footnotesize (6)} \\
$S_4$ & 0.1717 {\footnotesize (7)} & 0.3247 {\footnotesize (9)} & 0.3161 {\footnotesize (7)} & 0.4618 {\footnotesize (7)} \\
$S_5$ & 0.1913 {\footnotesize (1)} & 0.3985 {\footnotesize (1)} & 0.3880 {\footnotesize (1)} & 0.5498 {\footnotesize (1)} \\
$S_6$ & 0.1804 {\footnotesize (5)} & 0.3577 {\footnotesize (3)} & 0.3559 {\footnotesize (3)} & 0.5303 {\footnotesize (3)} \\
$S_7$ & 0.1913 {\footnotesize (1)} & 0.3374 {\footnotesize (6)} & 0.3347 {\footnotesize (6)} & 0.4541 {\footnotesize (8)} \\
$S_8$ & 0.1609 {\footnotesize (10)} & 0.2957 {\footnotesize (11)} & 0.2918 {\footnotesize (11)} & 0.4274 {\footnotesize (11)} \\
$S_9$ & 0.1413 {\footnotesize (13)} & 0.2677 {\footnotesize (15)} & 0.2422 {\footnotesize (15)} & 0.3630 {\footnotesize (14)} \\
$S_{10}$ & 0.1391 {\footnotesize (14)} & 0.3111 {\footnotesize (10)} & 0.2727 {\footnotesize (13)} & 0.3560 {\footnotesize (15)} \\
$S_{11}$ & 0.1370 {\footnotesize (15)} & 0.2753 {\footnotesize (14)} & 0.2691 {\footnotesize (14)} & 0.4103 {\footnotesize (12)} \\
$S_{12}$ & 0.1652 {\footnotesize (9)} & 0.2891 {\footnotesize (12)} & 0.3070 {\footnotesize (9)} & 0.5108 {\footnotesize (5)} \\
$S_{13}$ & 0.1587 {\footnotesize (11)} & 0.2859 {\footnotesize (13)} & 0.2798 {\footnotesize (12)} & 0.4305 {\footnotesize (10)} \\
$S_{14}$ & 0.1500 {\footnotesize (12)} & 0.3365 {\footnotesize (7)} & 0.2972 {\footnotesize (10)} & 0.3857 {\footnotesize (13)} \\
$S_{15}$ & 0.1717 {\footnotesize (6)} & 0.3271 {\footnotesize (8)} & 0.3102 {\footnotesize (8)} & 0.4349 {\footnotesize (9)} \\
\bottomrule
\end{tabular}
\end{table}

For each system, we computed standard IR effectiveness metrics using the reference qrels and those derived from individual judgments by both human and LLM annotators. Each qrels and metric produced a distinct system ranking. Although nDCG is often the primary metric discussed in recent studies \cite[e.g.,][]{rahmani2025judgingjudges, bencke2024trust, upadhyay2024largescale}, we expanded our evaluation to investigate whether the high ranking correlation observed in prior research is a general phenomenon or metric-dependent. Thus, we assessed system rankings using four distinct criteria:

\begin{itemize}
    \item \textbf{Precision at 10 (P@10) and Recall at 10 (R@10):} The first two metrics focus solely on the presence of relevant documents at a fixed cutoff. P@10 evaluates the fraction of the top-10 results that are relevant, while R@10 evaluates the fraction of all known relevant documents successfully retrieved within that rank. Although they are simple and intuitive metrics, they treat relevance as a binary attribute, do not account for the position of documents within the ranked list, and are highly sensitive to substantially incomplete relevance judgments \cite{retrieval_eval_with_incomplete_information}.

    \item \textbf{Normalized Discounted Cumulative Gain at 10 (nDCG@10):} It is a gain-based evaluation metric that accounts for both graded relevance and ranking position \cite{ndcg}. It rewards systems that retrieve highly relevant documents at higher ranks and is more robust to incomplete relevance judgments, being consistently adopted as a primary evaluation metric across some TREC (Text REtrieval Conference) tracks \cite{trec_dl_usa_ndcg}.

    \item \textbf{Mean Reciprocal Rank (MRR):} MRR considers only the rank of the first relevant document for each query \cite{manning2008introduction}. It measures how quickly a user finds a useful result and is valuable for evaluating tasks where a single highly ranked answer satisfies the user's information need.
\end{itemize}

Ranking correlation is measured using Kendall's $\tau$ and Spearman's $\rho$ coefficients. Spearman's $\rho$ is equivalent to applying Pearson correlation over the rank positions, capturing overall positional differences. Kendall's $\tau$ is based on pairwise comparisons and measures the degree of agreement between rankings by comparing the number of concordant and discordant pairs, i.e., how often pairs of systems are ordered consistently across rankings \cite{new_rank_correlation_coef}. Statistical uncertainty was assessed using BCa (bias-corrected and accelerated) bootstrap confidence intervals \cite{efron2016computer}. Note that eight of the fifteen systems are BM25 variants; thus, the correlations are computed over a set containing several closely related systems.

\subsection{Results}

The experimental results are structured into two parts. Section \ref{sec:results_assessment_query_doc_pairs} focuses on the query-document pair level, analyzing the agreement between LLMs and human annotators. Section \ref{sec:results_assessment_sys_ranking} investigates the downstream impact of LLM-generated qrels, evaluating whether they preserve the relative ranking of IR systems when compared to the reference judgments.

\subsubsection{Assessment of query-document pairs annotated by humans and LLMs -- \textbf{RO2}}
\label{sec:results_assessment_query_doc_pairs}

\begin{figure}[!htbp]
    \centering
    \includegraphics[width=0.9\linewidth]{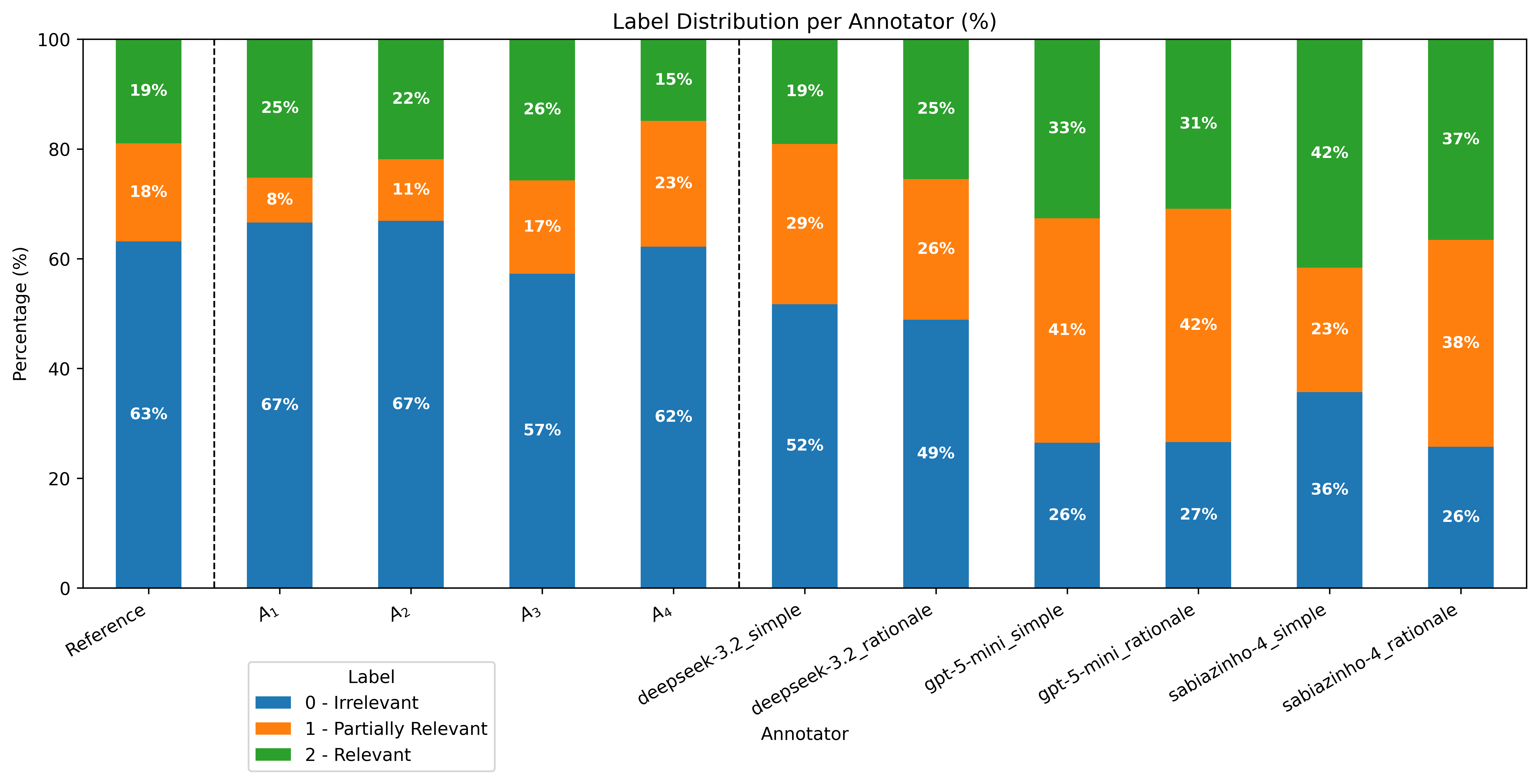}
    \caption{Label distribution (percentage) per annotator.}
    \label{fig:label_distribution_percentage_human_and_llm}
\end{figure}

Table \ref{tab:label_distribution_annotators_human_and_llm} reports the total number of labels assigned by each annotator (human and LLM), while Figure \ref{fig:label_distribution_percentage_human_and_llm} shows the corresponding percentage distribution. Both include the reference qrels and the labels assigned by all annotators.

All LLM annotators assigned fewer labels to the \texttt{irrelevant} (score 0) class than any human annotator and consequently assigned more labels to the \texttt{partially relevant} (score 1) and \texttt{relevant} (score 2) classes. We refer to this shift toward higher relevance scores as a positive bias. To measure this effect for each LLM annotator, we calculated the mean error (ME) and mean absolute error (MAE) between the LLM-assigned score and the average score of the human annotators at the query-document pair level. The results are shown in Table \ref{tab:bias_llm_human}. Note that the bias is calculated with respect to the human mean score rather than the reference qrels, as the latter is rounded after averaging.

\begin{table}[!htb]
\renewcommand{\arraystretch}{1.2}
\centering
\caption{Label distribution per annotator (absolute counts).}
\label{tab:label_distribution_annotators_human_and_llm}
\begin{tabular}{lrrrr}
\toprule
\textbf{Annotator} & \textbf{0 - irrelevant} & \textbf{1 - partially relevant} & \textbf{2 - relevant} & \textbf{Total} \\
\midrule
Reference qrels & 513  & 145  & 154  &  812 \\
\midrule
$A_1$ & 536 &  66 & 203 & 805 \\
$A_2$ & 477 &  80 & 156 & 713 \\
$A_3$ & 461 & 137 & 207 & 805 \\
$A_4$ & 451 & 166 & 108 & 725 \\
\midrule
deepseek-3.2\_simple     & 420 & 237 & 155 & 812 \\
deepseek-3.2\_rationale        & 397 & 208 & 207 & 812 \\
gpt-5-mini\_simple   & 215 & 332 & 265 & 812 \\
gpt-5-mini\_rationale      & 216 & 345 & 251 & 812 \\
sabiazinho-4\_simple   & 290 & 184 & 338 & 812 \\
sabiazinho-4\_rationale      & 209 & 306 & 297 & 812 \\
\bottomrule
\end{tabular}
\end{table}

The \texttt{deepseek-3.2} model had the lowest bias among the tested models. Using the \texttt{simple} prompt, the ME was only 0.08 points, but its MAE was 0.46, i.e., although the model diverges from the average human by nearly 0.5 points, it does so in an almost balanced way. In contrast, \texttt{gpt-5-mini} and \texttt{sabiazinho-4} showed a stronger positive bias.

Regarding the effect of prompting techniques on bias, there is no clear advantage in asking for a reason before giving the score. While the \texttt{rationale} prompt slightly improved the MAE for \texttt{gpt-5-mini} (from 0.60 to 0.58), it performed worse than the \texttt{simple} prompt for both \texttt{deepseek-3.2} (from 0.46 to 0.48) and \texttt{sabiazinho-4} (from 0.64 to 0.66). For all models, the differences between prompting techniques were marginal, with the MAE varying by at most 0.02 points across prompts.

Although the higher positive bias displayed by \texttt{sabiazinho-4} compared to \texttt{deepseek-3.2} may appear counterintuitive given the former's language specialization, such performance variations are not entirely unexpected. Evaluation benchmarks indicate that \texttt{deepseek-3.2} occasionally outperforms \texttt{sabiazinho-4} on specific tasks \cite{laitz2026sabia4technicalreport}. Moreover, pair-level agreement represents only one perspective on annotation quality. As discussed in the following section, \texttt{sabiazinho-4} often achieves stronger ranking correlations with the reference qrels than \texttt{deepseek-3.2}, despite exhibiting a higher positive bias at the query-document pair level.

\begin{table}[!htb]
\renewcommand{\arraystretch}{1.2}
\centering
\caption{Mean error (ME) and mean absolute error (MAE) of LLM annotators relative to human assessments, with corresponding standard deviation.}
\label{tab:bias_llm_human}
\begin{tabular}{lrrrr}
\toprule
 & \multicolumn{2}{c}{\textbf{Error}} & \multicolumn{2}{c}{\textbf{Absolute error}} \\
\cmidrule(lr){2-3} \cmidrule(lr){4-5}
\textbf{LLM annotator} & \textbf{ME} & \textbf{Std.} & \textbf{MAE} & \textbf{Std.} \\
\midrule
deepseek-3.2\_simple & 0.08 & 0.70 & 0.46 & 0.54 \\
deepseek-3.2\_rationale & 0.18 & 0.73 & 0.48 & 0.57 \\
gpt-5-mini\_simple & 0.47 & 0.70 & 0.60 & 0.58 \\
gpt-5-mini\_rationale & 0.45 & 0.68 & 0.58 & 0.57 \\
sabiazinho-4\_simple & 0.47 & 0.80 & 0.64 & 0.67 \\
sabiazinho-4\_rationale & 0.52 & 0.73 & 0.66 & 0.61 \\
\bottomrule
\end{tabular}
\end{table}

We also checked how frequently each LLM's scores fell within the range of labels assigned by human annotators. Table~\ref{tab:disagreement_by_model} reports the percentage of query-document pairs for which the LLM annotator agreed with at least one human judgment, stratified by category.

Within the 437 pairs with full consensus among human annotators, performance varied across relevance levels. Agreement was generally higher for pairs labeled as \texttt{relevant} (score 2) than for \texttt{irrelevant} pairs (score 0) for most LLM annotators. The \texttt{deepseek-3.2} model achieved the highest overall alignment, matching the human label in approximately 75\% of the cases, regardless of the prompt technique, largely driven by the high number of agreements in the \texttt{irrelevant} class.

For the 225 pairs exhibiting weak human disagreement (a one-point difference between annotators), all models agreed with at least one of the human assessments in at least 73\% of the cases. Agreement was particularly high for pairs with scores in the range 1-2, where several LLM annotators exceeded 90\% alignment with the range of human scores.

\begin{table}[!htb]
\renewcommand{\arraystretch}{1.2}
\centering
\caption{LLM agreement with human relevance intervals across subsets of query-document pairs defined by human disagreement levels. Human rows report the number of pairs in each category, while totals and percentages for LLM judges indicate agreement within those same pairs.}
\label{tab:disagreement_by_model}
\begin{tabular}{lcccc}
\toprule
\multicolumn{5}{c}{\textbf{Pairs with human consensus}} \\
\midrule
\textbf{Annotator} & \textbf{Score 0} & \textbf{Score 1} & \textbf{Score 2} & \textbf{Total} \\
\midrule
Human & 360 & 11 & 66 & 437 \\
deepseek-3.2\_simple & 278 (77\%) & 4 (36\%) & 47 (71\%) & 329 (75\%) \\
deepseek-3.2\_rationale & 273 (76\%) & 5 (45\%) & 54 (82\%) & 332 (76\%) \\
gpt-5-mini\_simple & 185 (51\%) & 6 (55\%) & 56 (85\%) & 247 (57\%) \\
gpt-5-mini\_rationale & 190 (53\%) & 5 (45\%) & 57 (86\%) & 252 (58\%) \\
sabiazinho-4\_simple & 222 (62\%) & 3 (27\%) & 57 (86\%) & 282 (65\%) \\
sabiazinho-4\_rationale & 173 (48\%) & 4 (36\%) & 56 (85\%) & 233 (53\%) \\
\midrule
\multicolumn{5}{c}{\textbf{Pairs with human weak disagreement}} \\
\midrule
\textbf{Annotator} & \textbf{Scores 0--1} & \textbf{Scores 1--2} & & \textbf{Total} \\
\midrule
Human & 135 & 90 & & 225 \\
deepseek-3.2\_simple & 118 (87\%) & 74 (82\%) & & 192 (85\%) \\
deepseek-3.2\_rationale & 109 (81\%) & 77 (86\%) & & 186 (83\%) \\
gpt-5-mini\_simple & 93 (69\%) & 87 (97\%) & & 180 (80\%) \\
gpt-5-mini\_rationale & 101 (75\%) & 90 (100\%) & & 191 (85\%) \\
sabiazinho-4\_simple & 80 (59\%) & 84 (93\%) & & 164 (73\%) \\
sabiazinho-4\_rationale & 86 (64\%) & 87 (97\%) & & 173 (77\%) \\
\midrule
\multicolumn{4}{l}{\textbf{Pairs with human strong disagreement}} & 150 \\
\bottomrule
\end{tabular}
\end{table}

\begin{figure}[!htbp]
    \centering
    \includegraphics[width=1\linewidth]{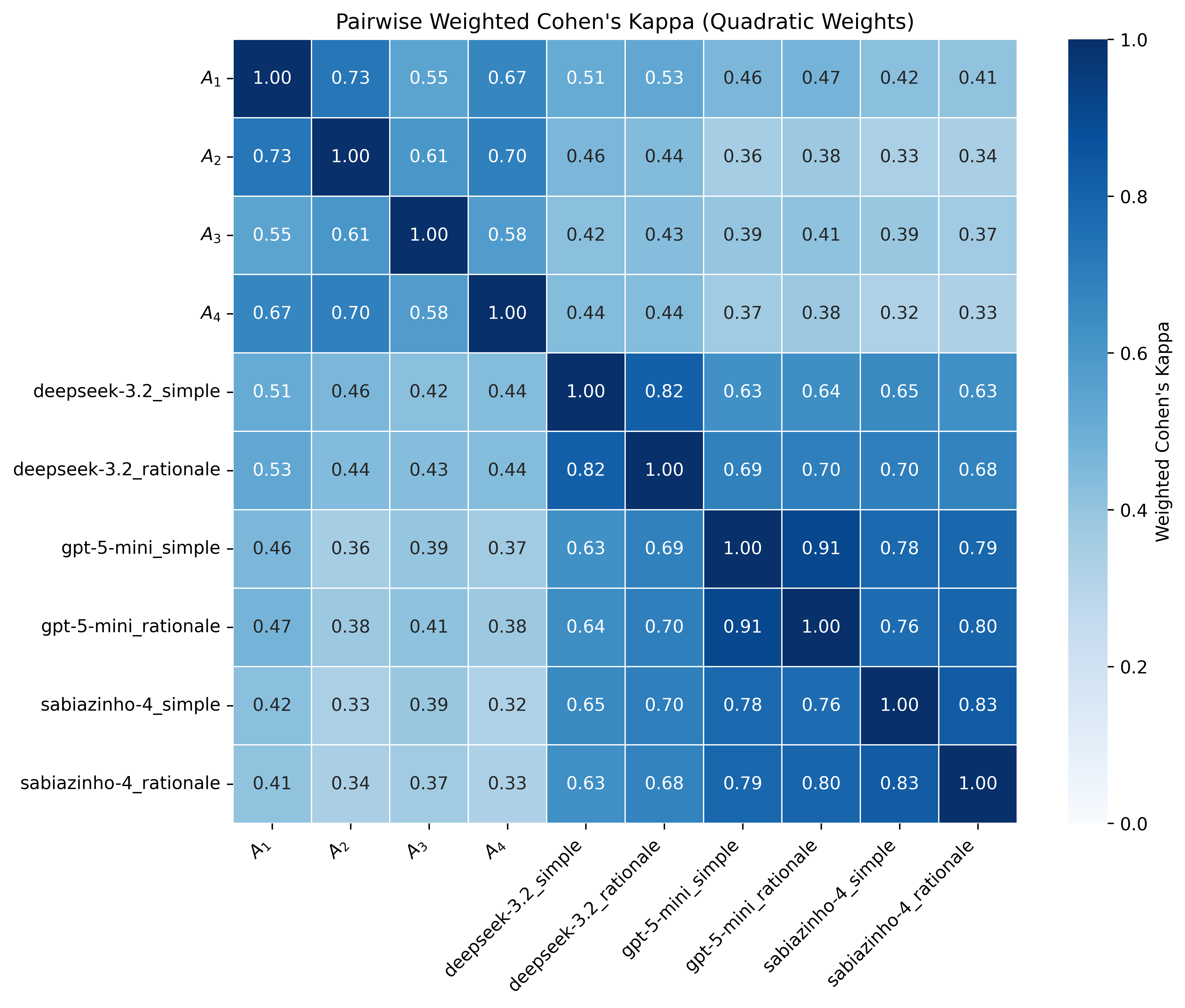}
    \caption{Pairwise weighted Cohen's $\kappa$ (quadratic) between human and LLM annotators.}
    \label{fig:pairwise_kappa_human_and_llm}
\end{figure}

Finally, Figure \ref{fig:pairwise_kappa_human_and_llm} shows the pairwise quadratic weighted Cohen's $\kappa$ across all annotators. The 4$\times$4 upper-left quadrant corresponds to the agreement among the human annotators, and the 6$\times$6 lower-right quadrant shows agreement among the LLM annotators.

Considering the conventional interpretation of Cohen's $\kappa$ shown in Table \ref{tab:kappa_interpretation}, human annotators exhibit moderate to substantial within-group agreement ($\kappa$ ranging from 0.55 to 0.73), whereas LLM annotators show substantial to almost perfect within-group agreement ($\kappa$ ranging from 0.63 to 0.91).

Within the LLM group, changing the prompt for a given model produces different label distributions (Table \ref{tab:label_distribution_annotators_human_and_llm} and Figure \ref{fig:label_distribution_percentage_human_and_llm}) but maintains high intra-model consistency. For example, agreement between the \texttt{simple} and \texttt{rationale} prompts, as measured by $\kappa$, was 0.82 for \texttt{deepseek-3.2}, 0.91 for \texttt{gpt-5-mini}, and 0.83 for \texttt{sabiazinho-4}.

However, it is important to examine the alignment among LLM and human annotators (the lower-left 4$\times$6 quadrant). Here, $\kappa$ ranged from 0.32 (\texttt{sabiazinho-4\_simple} and $A_4$) to 0.53 (\texttt{deepseek-3.2\_rationale} and $A_1$). While most LLM-human pairs achieved moderate agreement ($0.41 \le \kappa \le 0.60$), a few fell into the fair agreement range ($0.21 \le \kappa \le 0.40$). This low agreement may be partially explained by the positive bias of the LLMs, as shown in Table \ref{tab:bias_llm_human}. Note that the smaller the bias indicated for a given LLM annotator, the higher the corresponding $\kappa$.

Although agreement is low in some cases, the goal of judging query-document pairs is ultimately to produce qrels that can be used to rank IR systems.

\subsubsection{Assessment of the use of qrels generated by humans and by LLMs to rank IR systems -- \textbf{RO3}}
\label{sec:results_assessment_sys_ranking}

This analysis evaluates whether LLM-generated qrels preserve the relative ordering of the fifteen IR systems described in Section~\ref{sec:method}, despite the pair-level disagreements reported in Section~\ref{sec:results_assessment_query_doc_pairs}. We also analyze rankings produced by individual human annotators ($A_1$ to $A_4$).

\begin{figure}[!htbp]
    \centering
    \includegraphics[width=1\linewidth]{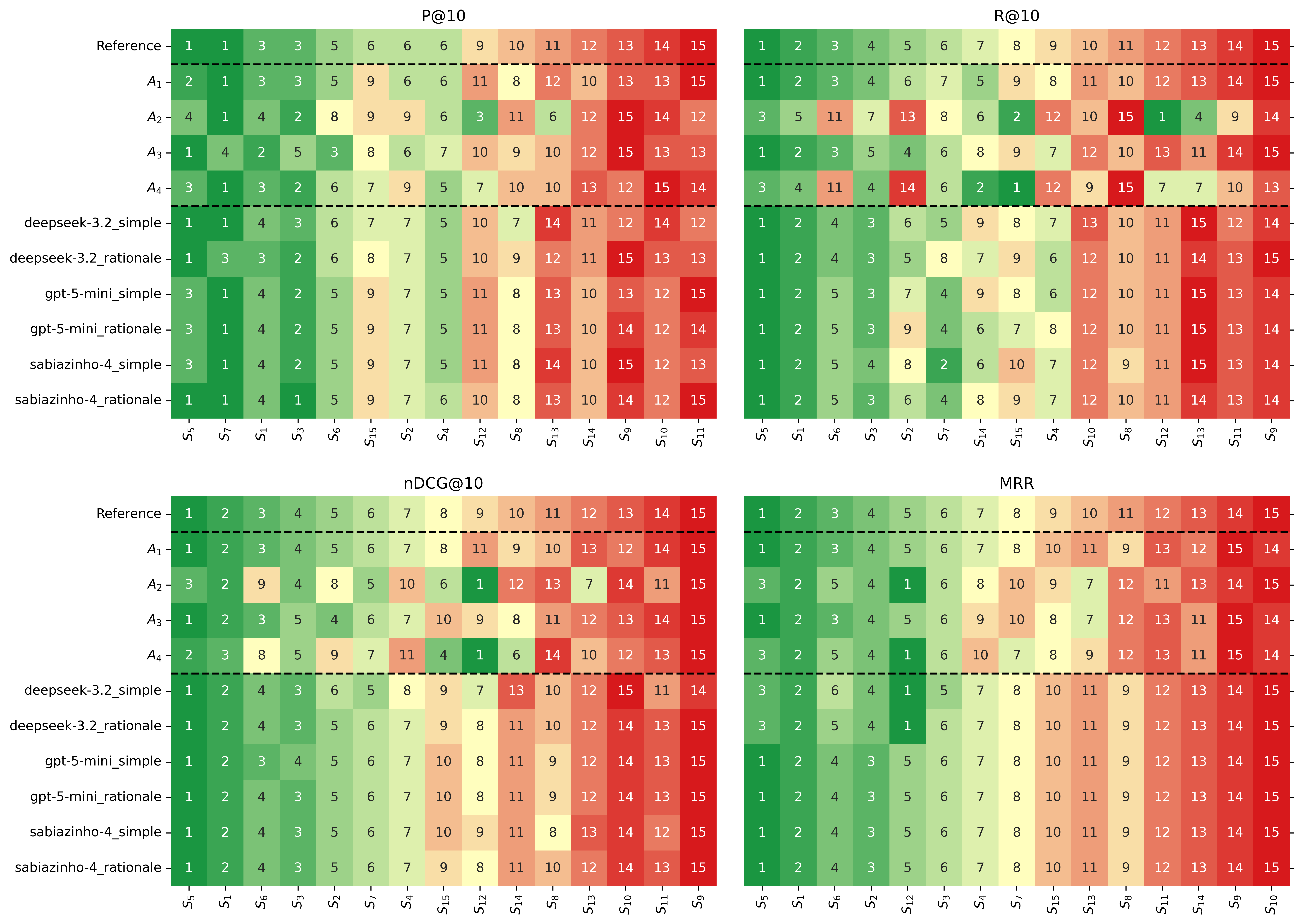}
    \caption{Heatmaps showing the ranking of retrieval systems across different qrels and evaluation metrics. Each cell shows the rank of a system (columns) under a given set of qrels (rows), represented both by color and numerically. Lower values indicate better performance (rank 1 = best), and equal numbers in the same row indicate a tie. Columns represent the fifteen IR systems and are ordered according to the reference ranking. Rows are grouped into reference, human, and LLM-generated qrels, with dashed lines separating these groups. Rankings are computed using P@10 (top-left), R@10 (top-right), nDCG@10 (bottom-left), and MRR (bottom-right). When alternative qrels are used for ranking, deviations from the reference ordering become visible as disruptions in the color gradient and in the numerical ordering of systems across rows. Larger deviations correspond to more substantial rank swaps.}
\label{fig:heatmap_rankings}
\end{figure}

Figure \ref{fig:heatmap_rankings} shows the rankings produced using the reference qrels and those derived from each human and LLM annotator. The rankings are represented as heatmaps, where rows correspond to different qrels and columns correspond to the fifteen IR systems. Rankings are encoded both numerically and by color, where lower values (rank 1) and greener tones indicate better performance. The first row shows the reference ranking, with systems ordered along the x-axis from best to worst according to the evaluation metric. Subsequent rows show the rankings generated by each annotator.

Deviations from the reference ranking are reflected as changes in both the color gradient and the numerical ordering of systems across rows. When a ranking matches the reference, the same color gradient is preserved. Small deviations appear as minor disruptions in the gradient, while larger deviations result in more pronounced color changes and reordering, indicating substantial rank swaps.

Although each qrels yields different absolute metric values, what matters is whether they preserve the relative ordering of the evaluated systems. That is, consistency in ranking is more important than agreement in raw scores. For transparency, \ref{sec:metrics_across_qrels} presents the same information shown in Figure \ref{fig:heatmap_rankings} using scatter plots, showing the actual metric values computed for each qrels.

To measure ranking stability, Table~\ref{tab:kendall_spearman} reports Kendall's $\tau$ and Spearman's $\rho$ correlation coefficients between the rankings generated using the reference qrels and those derived from qrels generated by individual human annotators and LLMs. Although there is no strict rule for interpreting these coefficients, Voorhees considers rankings with $\tau \ge 0.9$ to be functionally equivalent, and rankings with $\tau < 0.8$ to exhibit noticeable differences in general \cite{eval_by_highly_rel_doc}, a convention we adopt throughout our analysis.

\begin{table}[!htb]
\renewcommand{\arraystretch}{1.2}
\centering
\caption{Kendall's $\tau$ and Spearman's $\rho$ correlations between IR system rankings produced with each qrels set and the ranking obtained using the reference qrels across P@10, R@10, nDCG@10, and MRR.}
\label{tab:kendall_spearman}
\begin{tabular}{lcccccccc}
\toprule
 & \multicolumn{2}{c}{P@10} & \multicolumn{2}{c}{R@10} & \multicolumn{2}{c}{nDCG@10} & \multicolumn{2}{c}{MRR} \\
\cmidrule(lr){2-3} \cmidrule(lr){4-5} \cmidrule(lr){6-7} \cmidrule(lr){8-9}
\textbf{qrels set} & $\tau$ & $\rho$ & $\tau$ & $\rho$ & $\tau$ & $\rho$ & $\tau$ & $\rho$ \\
\midrule
$A_1$ & 0.90 & 0.97 & 0.92 & 0.98 & 0.94 & 0.99 & 0.92 & 0.98 \\
$A_2$ & 0.67 & 0.82 & 0.16 & 0.21 & 0.54 & 0.70 & 0.79 & 0.93 \\
$A_3$ & 0.85 & 0.95 & 0.89 & 0.97 & 0.92 & 0.98 & 0.85 & 0.95 \\
$A_4$ & 0.86 & 0.96 & 0.30 & 0.42 & 0.54 & 0.69 & 0.77 & 0.92 \\
\midrule
deepseek-3.2\_simple & 0.84 & 0.94 & 0.81 & 0.94 & 0.81 & 0.94 & 0.83 & 0.94 \\
deepseek-3.2\_rationale & 0.85 & 0.96 & 0.85 & 0.96 & 0.92 & 0.99 & 0.85 & 0.95 \\
gpt-5-mini\_simple & 0.83 & 0.94 & 0.79 & 0.93 & 0.92 & 0.98 & 0.94 & 0.99 \\
gpt-5-mini\_rationale & 0.83 & 0.94 & 0.81 & 0.93 & 0.90 & 0.97 & 0.94 & 0.99 \\
sabiazinho-4\_simple & 0.79 & 0.92 & 0.77 & 0.91 & 0.87 & 0.96 & 0.94 & 0.99 \\
sabiazinho-4\_rationale & 0.87 & 0.95 & 0.84 & 0.95 & 0.92 & 0.99 & 0.94 & 0.99 \\
\bottomrule
\end{tabular}
\end{table}

For the ranking generated by the P@10 metric, $0.79 \le \tau \le 0.87$ for the LLM annotators. From Voorhees' perspective, these rankings are not equivalent, and some annotators exhibited substantial position changes (e.g., several swaps within the top-4 positions). For the human annotators, the level of correlation was relatively high, except for $A_2$, where the system originally ranked third in the reference ranking dropped to the ninth position.

One possible explanation for these results is that P@10 is not a powerful discriminator among retrieval methods, as it only reflects the fraction of relevant documents among the top-10 retrieved \cite{retrieval_eval_with_incomplete_information}; thus, a single document no longer appearing in the top-10 list can change the metric by 10\% and destabilize the rankings. Furthermore, this issue is likely compounded by the limited size of the test collection. As Voorhees concluded, collections with a query count in the order of 50 are generally insufficient to achieve high confidence and stability when evaluating systems with P@10 \cite{voorhees2001philosophy, precision50queries}.

For the human annotators, sensitivity to missing judgments may be one contributing factor to ranking divergence, since annotators $A_2$ and $A_4$ evaluated 99 and 87 fewer query-document pairs than the reference qrels, respectively (Table~\ref{tab:label_distribution_annotators}). However, substantial ranking divergence was observed primarily for $A_2$. For the LLMs, the observed ranking instabilities are likely driven by model bias, leading some \texttt{irrelevant} documents to be judged as \texttt{relevant} or \texttt{partially relevant}.

A similar analysis applies to the R@10 metric. The rankings produced by LLM annotators are not equivalent to the reference ranking ($0.77 \le \tau \le 0.85$), with substantial rank changes for some systems. Among the human annotators, there was considerable variability: two annotators ($A_1$ and $A_3$) produced rankings very close to the reference, while the other two ($A_2$ and $A_4$) produced markedly different rankings, including a system moving up from the 11th to the 3rd position and the top-ranked system dropping to the 12th ($A_2$) or 8th position ($A_4$). This may be explained because R@10 behaves similarly to P@10. While both measure the proportion of relevant documents retrieved in the top-10 results, R@10 normalizes this count by the total number of relevant documents for each query. In the case of the \thisdataset dataset, there are approximately 6.5 relevant documents per query (Table \ref{tab:label_distribution_annotators_human_and_llm}), making R@10 even more sensitive than P@10 to missing judgments and small variations in the retrieved set.

For the rankings generated using nDCG@10, the LLM annotators using the \texttt{rationale} prompt achieved $0.90 \leq \tau \leq 0.92$, which, according to Voorhees, indicates that they are functionally equivalent to the reference ranking. For annotators using the \texttt{simple} prompt, there is greater variation across models ($0.81 \leq \tau \leq 0.92$). Among them, \texttt{gpt-5-mini\_simple} also reached the functional equivalence threshold ($\tau=0.92$), whereas \texttt{deepseek-3.2\_simple} and \texttt{sabiazinho-4\_simple} produced rankings that remained highly correlated with the reference ranking. The first seven positions across the rankings are highly stable, with only minor reordering among closely ranked systems (limited to one-position shifts). Among the human annotators, the rankings derived from $A_1$ and $A_3$ are highly correlated with the reference ranking, whereas those derived from $A_2$ and $A_4$ show lower correlation.

For the MRR metric, the rankings generated by LLM annotators using the \texttt{gpt-5-mini} and \texttt{sabiazinho-4} models are equivalent to the reference ranking ($\tau = 0.94$), with only minor deviations (a single swap between the 3rd and 4th positions and the 9th-place system shifting to 11th). In contrast, those produced by the \texttt{deepseek-3.2} model showed a high, albeit slightly lower, correlation ($0.83 \leq \tau \leq 0.85$). Among human annotators, only the ranking produced by $A_1$ is equivalent to the reference ($\tau = 0.92$). The rankings produced using MRR exhibit behavior similar to that of nDCG@10, but with lower variability, likely because MRR relies on binary relevance rather than graded relevance and considers only the position of the first relevant document, rather than the overall ordering of the ranked list.

Across the four evaluated metrics, the rankings generated by LLM annotators are, in most cases, equivalent to the reference rankings for nDCG@10 and MRR. The \texttt{rationale} prompt generally performs better, as it consistently yields $\tau \ge 0.9$ for these metrics (except for MRR and \texttt{deepseek-3.2\_rationale}). For nDCG@10, the \texttt{simple} prompt achieved better results than \texttt{rationale} only for the \texttt{gpt-5-mini} model. For MRR, the \texttt{gpt-5-mini} and \texttt{sabiazinho-4} models produced identical rankings regardless of the prompt used. The results also show that LLMs cannot reliably reproduce the system rankings generated with P@10 and R@10, as these metrics lead to significant deviations from the reference ranking (according to the functional equivalence criterion proposed by Voorhees \cite{eval_by_highly_rel_doc}).

It is also informative to compare the rankings generated by individual human annotators with the reference ranking. As shown in Table~\ref{tab:kendall_spearman}, relying on a single human assessor introduces substantial variability due to differences in how annotators interpret and apply relevance criteria. On the other hand, rankings generated by LLMs tend to exhibit lower variability among themselves and, in some cases, achieve higher correlation than those produced by individual human annotators. For instance, for MRR, the Kendall's $\tau$ values for annotators $A_2$, $A_3$, and $A_4$ are lower than or equal to those obtained with all LLM-based rankings except \texttt{deepseek-3.2\_simple}.

\begin{figure}[!htbp]
    \centering
    \includegraphics[width=1\linewidth]{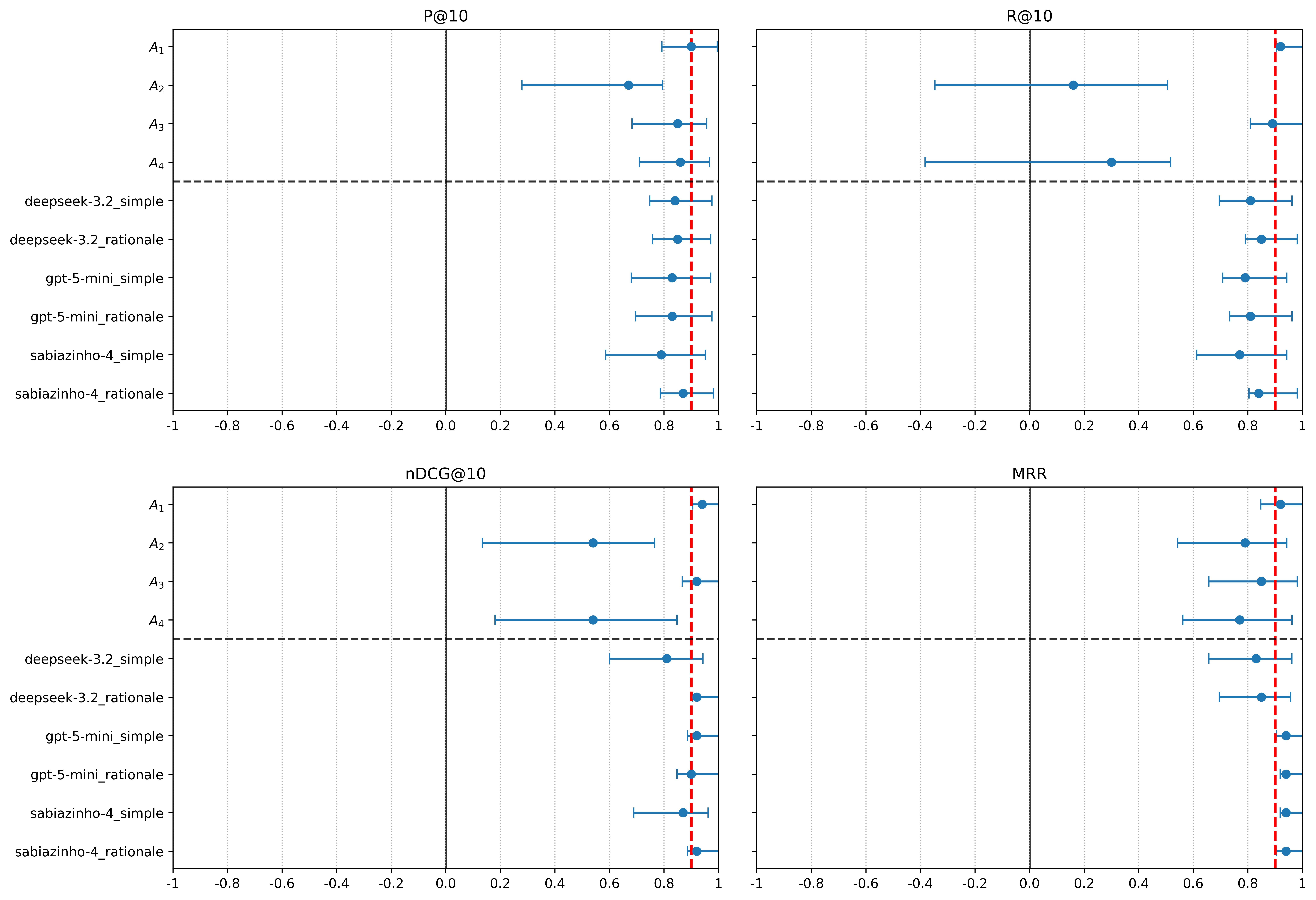}
    \caption{Bootstrap analysis of ranking agreement. Circles represent the observed Kendall's $\tau$ values reported in Table~\ref{tab:kendall_spearman}, and error bars indicate 90\% BCa bootstrap confidence intervals. For each bootstrap sample, queries were resampled with replacement and Kendall's $\tau$ was recomputed between the ranking produced by the reference qrels and by the human and LLM annotators using the same resampled queries. The dashed red line indicates the threshold $\tau = 0.9$, which Voorhees considers indicative of functionally equivalent rankings \cite{eval_by_highly_rel_doc}.}
\label{fig:bootstrap_ic_90}
\end{figure}

To assess the robustness of the ranking correlations under different query samples, we performed a BCa bootstrap analysis with 10,000 resamples of the 46 queries, sampled with replacement. Figure~\ref{fig:bootstrap_ic_90} presents the results. The circles represent the observed Kendall's $\tau$ values reported in Table~\ref{tab:kendall_spearman}, while the error bars indicate 90\% confidence intervals. The bootstrap analysis supports our previous ranking correlation analysis.

In several scenarios, the confidence intervals associated with the LLM annotators were narrower than those of individual human annotators. This is noticeable for $A_2$ across all evaluation metrics, $A_3$ for MRR, $A_4$ for R@10, nDCG@10, and MRR, and even for $A_1$ for MRR when compared against the \texttt{gpt-5-mini} and \texttt{sabiazinho-4} models. These narrower intervals might indicate that, for certain query subsets, individual human judgments deviate more from the collective human average (reference ranking) than the judgments generated by the LLMs.

The confidence intervals for the LLM annotators generally do not lie entirely above the threshold of $\tau \ge 0.9$. However, there are some exceptions: for MRR, the confidence intervals for the \texttt{gpt-5-mini} and \texttt{sabiazinho-4} models were entirely above this threshold, as was the interval for \texttt{deepseek-3.2\_rationale} on nDCG@10. Regarding nDCG@10 and the \texttt{rationale} prompts, the observed $\tau$ exceeded 0.9, and the confidence intervals remained almost entirely above this threshold.

For the human annotators, only $A_1$ achieved confidence intervals entirely above the threshold, and only for R@10 and nDCG@10. Particularly noteworthy are the confidence intervals for $A_2$ and $A_4$ under R@10, which extend into $\tau < 0$. This result is likely due to the sensitivity of R@10 to missing judgments and to the relatively small number of relevant documents per query in the \thisdataset collection, as discussed earlier.

\section{Discussion of the results, implications, and limitations of the study}

\subsection{Discussion of the results}

In this study, we introduced a new dataset (\thisdataset) for legal IR and used it to investigate the reliability of LLMs for relevance assessment in Portuguese. Our analysis of human annotations showed that disagreement among domain experts was common. While more than half of the query-document pairs achieved full consensus among the human annotators, a substantial proportion of the annotated pairs exhibited weak disagreement (28\%, indicating a difference of only one relevance level) or strong disagreement (18\%, where scores of both 0 and 2 were assigned to the same pair). However, part of this observed variability was due to legitimate differences in semantic interpretation rather than annotation noise or lack of expertise, indicating the inherent subjectivity of this task.

When comparing LLM and human annotators, the experiment was consistent with patterns previously reported in the literature. LLMs tended to assign higher relevance scores than human annotators, resulting in a positive bias. Although the magnitude of this bias varied substantially across models, no significant differences were observed between the two prompting techniques tested (\texttt{simple} and \texttt{rationale}) at the pair-annotation level. Among the evaluated models, \texttt{deepseek-3.2} achieved the lowest MAE.

Despite this bias, it is important to evaluate the end-to-end task, as absolute scoring differences may affect the final result in different ways. In the context of information retrieval, the primary objective of relevance assessment is to reliably rank retrieval systems. By testing four evaluation metrics for ranking generation (P@10, R@10, nDCG@10, and MRR), we observed that rankings based on LLM-generated qrels remained highly correlated with the reference ranking when evaluated using nDCG@10 and MRR. However, this ranking consistency did not hold for P@10 and R@10, which showed substantial deviations. Additionally, while the prompting technique had little effect on pair-level bias, the \texttt{rationale} prompt proved superior for system-level evaluation, consistently producing rankings that were functionally equivalent to the reference ranking for nDCG@10 and MRR. Furthermore, unlike the pair-level analysis, in which \texttt{deepseek-3.2} achieved the lowest MAE, the strongest end-to-end results were obtained by \texttt{gpt-5-mini} and \texttt{sabiazinho-4}, indicating that lower annotation error does not necessarily translate into better preservation of system rankings.

While recent literature validating LLM-as-a-judge often relies heavily on nDCG, some studies have evaluated other metrics. For instance, Upadhyay et al evaluated Recall at 100 (R@100) and reported high ranking consistency \cite{upadhyay2024largescale}. The divergence observed in our study is likely due to the use of a much shallower cutoff (@10 instead of @100) combined with the scarcity of relevant documents per query. In specialized domains, such as the legal corpus of \thisdataset, it is common to have few documents that effectively answer a query, in contrast to general web datasets. For \thisdataset, there are only approximately 6.5 relevant documents per query on average (Table \ref{tab:label_distribution_annotators_human_and_llm}). Consequently, R@10 becomes substantially more sensitive than R@100; a single relevant document dropping out of the top-10 list drastically impacts the metric score and destabilizes the resulting system rankings. It is worth noting that, from a user experience perspective, shallow cutoffs like @10 are critical, as they represent the first page of search results where users concentrate almost all their attention.

\subsection{Practical implications}

Based on these findings, this study has the following practical implications:

\begin{itemize}
    \item The positive bias observed in LLM-based relevance assessment may extend to other domains. Accordingly, LLMs should be evaluated on the end-to-end task for which they are intended, rather than only at the annotation step.

    \item The reliability of LLM-generated judgments for system evaluation is sensitive to the chosen evaluation metric. In specialized domains, where relevant documents are often scarce, the results suggest that LLM-as-a-judge may be used to compare IR systems when nDCG@10 and MRR are adopted (rank-aware metrics), but it is inadequate for comparing rankings based on P@10 or R@10.
    
    \item For the task of generating relevance judgments intended for system ranking, the \texttt{rationale} prompt should be preferred. Even though it did not reduce the absolute scoring bias at the pair level, it yielded more stable and accurate system rankings than the \texttt{simple} prompt.

    \item While exhaustive human annotation by multiple domain experts remains the gold standard, natural inter-annotator variability suggests a practical trade-off: when only a single human annotator is available due to budget or time constraints, LLM-generated judgments may provide a stable baseline for ranking IR systems using nDCG@10 or MRR under the conditions examined in this study.
\end{itemize}

\subsection{Limitations of the study}

While this study provides useful insights, some limitations should be acknowledged.

First, the dataset contains 46 queries. While larger query sets are generally preferable, this size is comparable to that of some established IR evaluation campaigns. For example, the TREC-COVID and TREC Deep Learning 2020 tracks provided 50 and 45 annotated queries, respectively, for the document retrieval task \cite{treccovid,trecdl2020}. Moreover, previous studies have shown that collections containing on the order of 50 queries can support reliable comparative evaluations \cite{zobel1998reliable,sanderson2010test,measurestability}.

Second, the experiment using LLM-as-a-Judge for relevance assessment was conducted solely on the \thisdataset dataset, and its results may not generalize to other collections. The high-ranking consistency observed with nDCG@10 and MRR aligns with recent literature \cite[e.g.,][]{thomas2024searcherpref, rahmani2025judgingjudges, bencke2024trust}. On the other hand, although the poor performance on P@10 and R@10 suggests that these metrics may be less suitable for LLM-as-a-judge frameworks, this divergence could also be due to the number of queries of the \thisdataset dataset, since these metrics are naturally more volatile when computed over fewer queries. While the collection size is comparable to that of established IR evaluation campaigns, metrics such as P@10 and R@10 are inherently more volatile when computed over fewer queries \cite{voorhees2001philosophy}.

Third, due to computational and API costs, we limited the experiment to six model-prompt combinations. Thus, the results reflect the behavior of a specific subset of state-of-the-art LLMs, and evaluating a broader range of models could lead to different ranking correlations and bias patterns. Furthermore, all evaluated models were accessed through proprietary APIs, whose implementations may change over time without public notice. Although we reported the model versions used and released the generated relevance judgments, future model updates may affect reproducibility. Nevertheless, we consider the reproducibility of the experimental methodology more important than the long-term availability of any particular model. Because the dataset also includes the individual human annotations, future studies can extend our evaluation using newer LLMs.

Finally, we evaluated only two prompting techniques (\texttt{simple} and \texttt{rationale}), which produced small differences in pair-level bias (measured by MAE). Exploring a broader range of prompting strategies was outside the scope of this work due to computational and API costs, but prior research in Portuguese Natural Language Inference (NLI) has shown that prompt variations can substantially affect task performance \cite{publichearingbr}. Therefore, different prompts could lead to different bias patterns and system rankings, suggesting that prompt design may be a useful mechanism for mitigating model bias.

\section{Conclusion}

This paper introduced \thisdataset, a new Brazilian Portuguese IR test collection with 14,469 legal documents, 46 queries, and graded relevance judgments produced by four domain experts. The corpus consists of normative acts issued by the Brazilian Federal Court of Accounts and provides a realistic benchmark for legal information retrieval. We released the corpus, queries, aggregated qrels, and anonymized individual annotations.

To investigate whether LLMs could serve as scalable relevance assessors, we conducted an experiment comparing judgments generated by three LLMs (\texttt{deepseek-3.2}, \texttt{gpt-5-mini}, and \texttt{sabiazinho-4}) against human annotations to rank fifteen IR systems. The results with the \thisdataset dataset demonstrated a clear metric-dependent behavior: although LLMs consistently exhibited a positive scoring bias at the query-document pair level, they produced system rankings that were highly consistent with the reference rankings when evaluated using nDCG@10 and MRR (rank-aware metrics). However, this level of agreement was not observed for P@10 and R@10.

Future work should focus on building test collections for other legal databases. The legal domain is broad and highly heterogeneous, covering not only normative acts but also jurisprudence, contracts, legal opinions, and administrative decisions. Advancements in this area are particularly important, as many applications in this domain involve a preliminary IR stage (for instance, legal search systems and case law recommendation tools). Additionally, further research should continue to explore the viability of LLM-as-a-judge as an alternative to human annotators in specialized domains, particularly by addressing the limitations identified in this study. Given our results, evaluations of automated relevance assessments should consider a broader range of IR metrics to fully understand the limitations and generalizability of LLM-generated judgments.


\section*{Declaration of generative AI and AI-assisted technologies in the manuscript preparation process}

During the preparation of this work, the authors used AI-assisted tools in order to improve the language and readability. After using these tools, the authors reviewed and edited the content as needed and take full responsibility for the content of the published article.

\bibliography{bibliography}

\appendix
\renewcommand{\thesection}{Appendix \Alph{section}}
\renewcommand{\thetable}{A.\arabic{table}}
\setcounter{table}{0}

\section{Queries}
\label{sec:anexo_queries}

\renewcommand{\arraystretch}{1.3}
\begin{longtable}{|p{0.47\linewidth}|p{0.47\linewidth}|}
    \caption{Original Portuguese queries in \thisdataset and their English translations. Queries are grouped into categories for descriptive purposes only.}
    \label{tab:queries}\\
    \hline
    \textbf{Query (Portuguese)} & \textbf{Translation (English)} \\ \hline
    \endfirsthead
    
    \multicolumn{2}{c}%
    {{\tablename\ \thetable{} -- continued from previous page}} \\
    \hline
    \textbf{Query (Portuguese)} & \textbf{Translation (English)} \\ \hline
    \endhead
    
    \hline
    \multicolumn{2}{|r|}{{Continued on next page}} \\ \hline
    \endfoot
    
    \hline
    \endlastfoot

    \multicolumn{2}{|c|}{{\textit{Administration and governance}}} \\ \hline
    regimento interno & rules of procedure\\
    horário de funcionamento do tcu & tcu opening hours\\
    utilização e indenização de despesas relacionadas aos serviços e aos dispositivos de telecomunicação & use and reimbursement of expenses related to telecommunication services and devices\\
    uso de garagem pernoite & overnight garage use\\
    código de ética & code of ethics\\
    quais as classificações de confidencialidade previstas para um documento & what confidentiality classifications are provided for a document\\
    constituição organização e tramitação de processos e documentos relativos à área administrativa & creation, organization, and processing of processes and documents related to the administrative area\\
    competências do Auditor-Chefe ao Auditor-Chefe Adjunto & responsibilities delegated by the Chief Auditor to the Deputy Chief Auditor\\
    diarias emissão de passagens & per diem and ticket issuance\\
    estrutura segecex & Segecex organizational structure\\
    proteção de dados pessoais & personal data protection\\
    manual de gestão de riscos do tcu & tcu risk management manual\\
    organização de processos de controle externo & organization of external control processes\\
    fase recursal dos auditores diretores e secretários que atuaram na fase de instrução originária & appeal phase of auditors, directors, and secretaries who acted in the first instruction phase\\
    Estabelece procedimentos para constituição organização e tramitação de processos e documentos relativos à área de controle externo & establishes procedures for creation, organization, and processing of processes and documents related to the external control area\\ \hline

    \multicolumn{2}{|c|}{{\textit{Staff regulations and human resources}}} \\ \hline
    trabalho fora das dependências do tcu & work outside tcu facilities \\
    quantos dias pode ter um período de licença capacitação & how many days can a professional development leave period have\\
    licença capacitação & professional development leave\\   
    qual a idade limite para a assistência escolar & what is the age limit for educational assistance\\
    jornada de trabalho no teletrabalho & workload in remote work\\
    autorização de acesso à declaração de bens e rendas & authorization to access income tax return\\
    limite de tempo para tirar licença para tratar de interesses particulares & time limit for leave to deal with private matters\\   
    dever do servidor no teletrabalho & civil servant duties in remote work\\
    feriados 2024 & 2024 holidays\\
    aposentadoria de servidores indenização & compensation in civil servants' retirement\\
    principais diretrizes de normas de auditoria do operacional do tribunal de contas da união & main guidelines of operational audit standards of the Brazilian Federal Court of Accounts\\
    aposentadoria por incapacidade permanente & retirement due to permanent disability\\
    "Dispõe sobre a assistência à saúde dos servidores ativos e inativos \& seus dependentes e pensionistas civis do TCU" & "Provides for health assistance for active and retired civil servants and their dependents and civil pensioners of TCU"\\
    RESOLUÇÃO GRATIFICAÇÃO DESEMPENHO & PERFORMANCE BONUS RESOLUTION\\
    progressão funcional & career progression\\
    Comissão de Avaliação de Desempenho dos Servidores do Tribunal de Contas da União & Performance Evaluation Committee of the Brazilian Federal Court of Accounts' civil servants\\ \hline

    \multicolumn{2}{|c|}{{\textit{Audit and external control}}} \\ \hline
    compra de pequeno valor & low-value purchase\\
    manual operacional & operational manual\\
    A empresa ainda não possui atestado de capacidade técnica pode-se utilizar atestado do responsável técnico & if the company does not yet have a technical capacity certificate, can the technical manager's certificate be used\\
    ausencia de custos unitários na proposta & absence of unit costs in the proposal\\
    prestação contas dirigente máximo & accountability report by the top manager\\
    atualização monetária prestação contas & monetary update accountability report\\
    valor máximo da multa a que se refere o caput do art. 58 2023 & maximum fine value referred to in the caput of art. 58 2023\\
    programa nacional de desestatização & national privatization program\\
    roteiro de auditoria de obras públicas & public works audit checklist\\
    manual de recurso & appeals manual\\
    monitoramento no próprio processo desestatização & monitoring within the privatization process itself\\
    questinonario de auditoria sobre a transparencia nos conselho de fisicalização & audit questionnaire on transparency in oversight councils\\
    manual de denúncia & complaint manual\\
    estratégia global auditoria & global audit strategy\\
    manual de auditoria de obras & public works auditing manual\\
\end{longtable}

\section{Prompts}
\label{sec:prompts}

Figures \ref{fig:prompt_simple} and \ref{fig:prompt_rationale} show the original prompts used in this paper, while Figures \ref{fig:prompt_simple_en} and \ref{fig:prompt_rationale_en} show their translations.

\begin{figure}[!htb]
\begin{tcolorbox}[
    left=2pt,
    right=2pt,
    top=2pt,
    bottom=2pt,
    halign=flush left
]
\footnotesize
\begin{verbatim}
[System]
Você é um avaliador de relevância de normas do Tribunal de Contas da União (TCU).

Sua tarefa é avaliar a relevância de um DOCUMENTO em relação a uma CONSULTA (QUERY), atribuindo um score de
relevância em {0, 1, 2}, conforme as regras abaixo.

### REGRAS PARA O SCORE ###:
2 - RELEVANTE: O documento responde diretamente à consulta e contém a informação central solicitada. O conteúdo
é suficiente para permitir ao usuário atingir o objetivo da busca. O documento não será considerado relevante
para a consulta se apenas mencionar termos semelhantes, mas não tratar efetivamente do objeto da consulta.

1 - POUCO RELEVANTE: O documento menciona o tema da consulta ou trata de assunto relacionado, mas de forma
parcial ou insuficiente para atender plenamente à necessidade informacional.

0 - IRRELEVANTE: O documento não trata do assunto da consulta ou aborda tema distinto/incompatível.
### ###

A sua resposta deverá ser apenas um JSON válido, sem qualquer texto antes ou depois, e deve conter apenas o
atributo "score".

### REGRAS PARA O JSON ###
"score": O valor do score (0, 1 ou 2) para o grau de relevância estimado.
### ###

[User]
Avalie a relevância do documento para a query.

QUERY: {QUERY}

DOCUMENTO:
{DOCUMENTO}
\end{verbatim}
\end{tcolorbox}

\caption{Simple relevance-judment prompt (Portuguese).}
\label{fig:prompt_simple}
\end{figure}
\begin{figure}[!htb]
\begin{tcolorbox}[
    left=2pt,
    right=2pt,
    top=2pt,
    bottom=2pt,
    halign=flush left
]
\footnotesize
\begin{verbatim}
[System]
Você é um avaliador de relevância de normas do Tribunal de Contas da União (TCU).

Sua tarefa é avaliar a relevância de um DOCUMENTO em relação a uma CONSULTA (QUERY), atribuindo um score de
relevância em {0, 1, 2}, conforme as regras abaixo.

### REGRAS PARA O SCORE ###:
2 - RELEVANTE: O documento responde diretamente à consulta e contém a informação central solicitada. O conteúdo
é suficiente para permitir ao usuário atingir o objetivo da busca. O documento não será considerado relevante
para a consulta se apenas mencionar termos semelhantes, mas não tratar efetivamente do objeto da consulta.

1 - POUCO RELEVANTE: O documento menciona o tema da consulta ou trata de assunto relacionado, mas de forma
parcial ou insuficiente para atender plenamente à necessidade informacional.

0 - IRRELEVANTE: O documento não trata do assunto da consulta ou aborda tema distinto/incompatível.
### ###

A sua resposta deverá ser apenas um JSON válido, sem qualquer texto antes ou depois, e deve conter apenas os
atributos "justificativa" e "score", nessa ordem.

### REGRAS PARA O JSON ###
"justificativa": Texto explicando se o documento é IRRELEVANTE, POUCO RELEVANTE ou RELEVANTE para a consulta.

"score": O valor do score (0, 1 ou 2) para o grau de relevância estimado.
### ###

[User]
Avalie a relevância do documento para a query.

QUERY: {QUERY}

DOCUMENTO:
{DOCUMENTO}
\end{verbatim}
\end{tcolorbox}

\caption{Rationale relevance-judgment prompt (Portuguese).}
\label{fig:prompt_rationale}
\end{figure}

\begin{figure}[!htb]
\begin{tcolorbox}[
    left=2pt,
    right=2pt,
    top=2pt,
    bottom=2pt,
    halign=flush left
]
\footnotesize
\begin{verbatim}
[System]
You are a relevance assessor of normative acts from the Brazilian Federal Court of Accounts (TCU).

Your task is to evaluate the relevance of a DOCUMENT with respect to a QUERY by assigning a relevance score in
{0, 1, 2}, according to the rules below.

### SCORING RULES ###:
2 - RELEVANT: The document directly answers the query and contains the core information requested. The content
is sufficient for the user to achieve their search objective. A document will not be considered relevant if it
only mentions similar terms without actually addressing the subject of the query.

1 - PARTIALLY RELEVANT: The document mentions the topic of the query or addresses a related subject, but in a
partial or insufficient manner to fully meet the information need.

0 - IRRELEVANT: The document does not address the subject of the query or deals with a different/incompatible
topic.
### ###

Your response must be a valid JSON only, with no text before or after, and must contain only the attribute
"score".

### JSON RULES ###
"score": The score value (0, 1, or 2) for the estimated relevance.
### ###

[User]
Assess the relevance of the document to the query.

QUERY: {QUERY}

DOCUMENT:
{DOCUMENT}
\end{verbatim}
\end{tcolorbox}

\caption{Simple relevance-judment prompt (English).}
\label{fig:prompt_simple_en}
\end{figure}
\begin{figure}[!htb]
\begin{tcolorbox}[
    left=2pt,
    right=2pt,
    top=2pt,
    bottom=2pt,
    halign=flush left
]
\footnotesize
\begin{verbatim}
[System]
You are a relevance assessor of normative acts from the Brazilian Federal Court of Accounts (TCU).

Your task is to evaluate the relevance of a DOCUMENT with respect to a QUERY by assigning a relevance score in
{0, 1, 2}, according to the rules below.

### SCORING RULES ###
2 - RELEVANT: The document directly answers the query and contains the core information requested. The content
is sufficient for the user to achieve their search objective. A document will not be considered relevant if it
only mentions similar terms without actually addressing the subject of the query.

1 - PARTIALLY RELEVANT: The document mentions the topic of the query or addresses a related subject, but in a
partial or insufficient manner to fully meet the information need.

0 - IRRELEVANT: The document does not address the subject of the query or deals with a different/incompatible
topic.
### ###

Your response must be a valid JSON only, with no text before or after, and must contain only the attributes
"reasoning" and "score", in this order.

### JSON RULES ###
"reasoning": Text explaining whether the document is IRRELEVANT, PARTIALLY RELEVANT, or RELEVANT to the query.

"score": The score value (0, 1, or 2) for the estimated relevance.
### ###

[User]
Assess the relevance of the document to the query.

QUERY: {QUERY}

DOCUMENT:
{DOCUMENT}
\end{verbatim}
\end{tcolorbox}
\caption{Rationale relevance-judgment prompt (English).}
\label{fig:prompt_rationale_en}
\end{figure}

\FloatBarrier

\section{Evaluation metrics for across qrels}
\label{sec:metrics_across_qrels}

\begin{figure}[!htbp]
    \centering
    \includegraphics[width=1\linewidth]{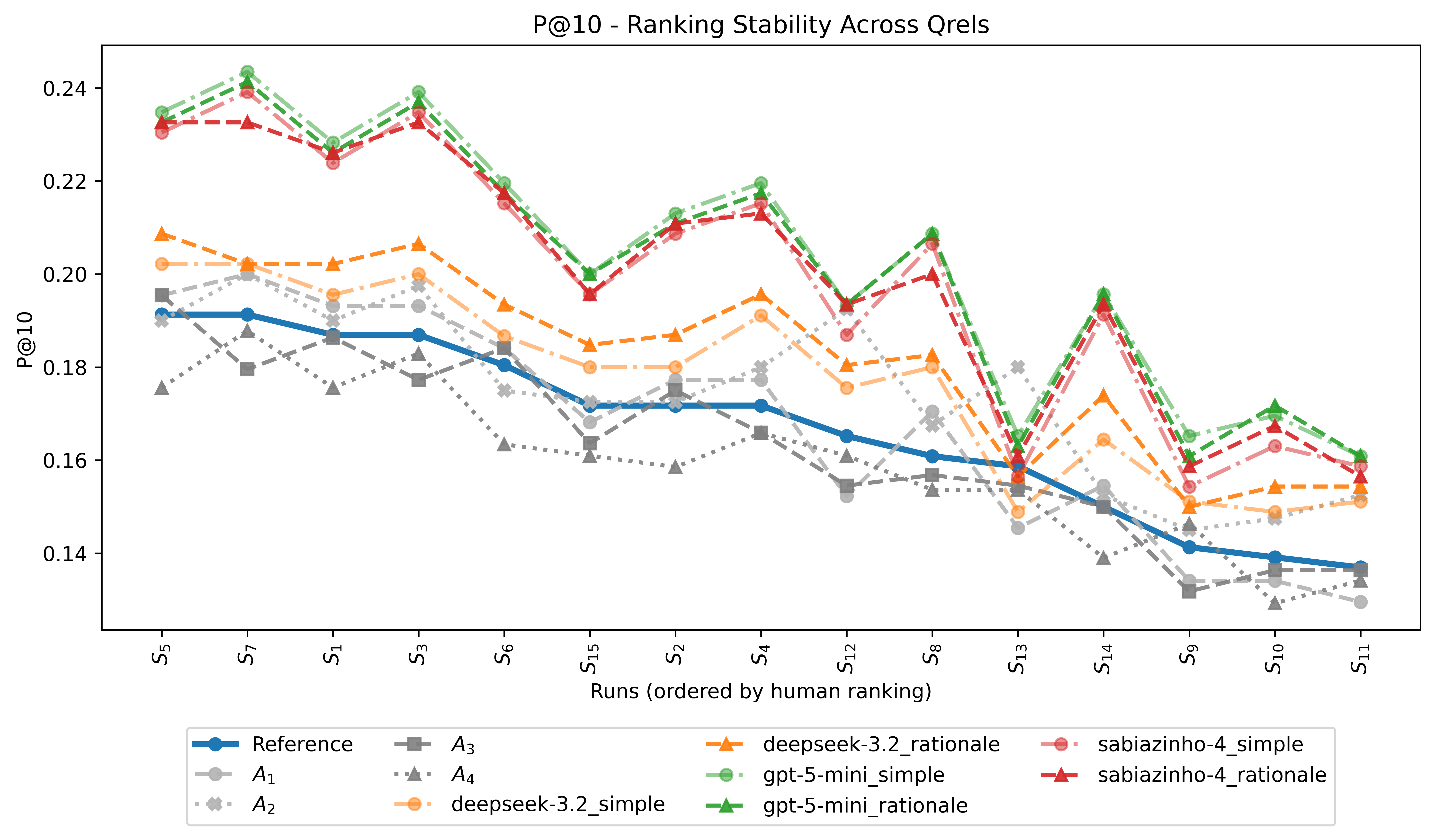}
    \caption{P@10 for each IR system computed with different qrels. Systems are ordered according to the ranking produced by the reference qrels.}
\label{fig:ranking_across_qrels_P_10}
\end{figure}

\begin{figure}[!htbp]
    \centering
    \includegraphics[width=1\linewidth]{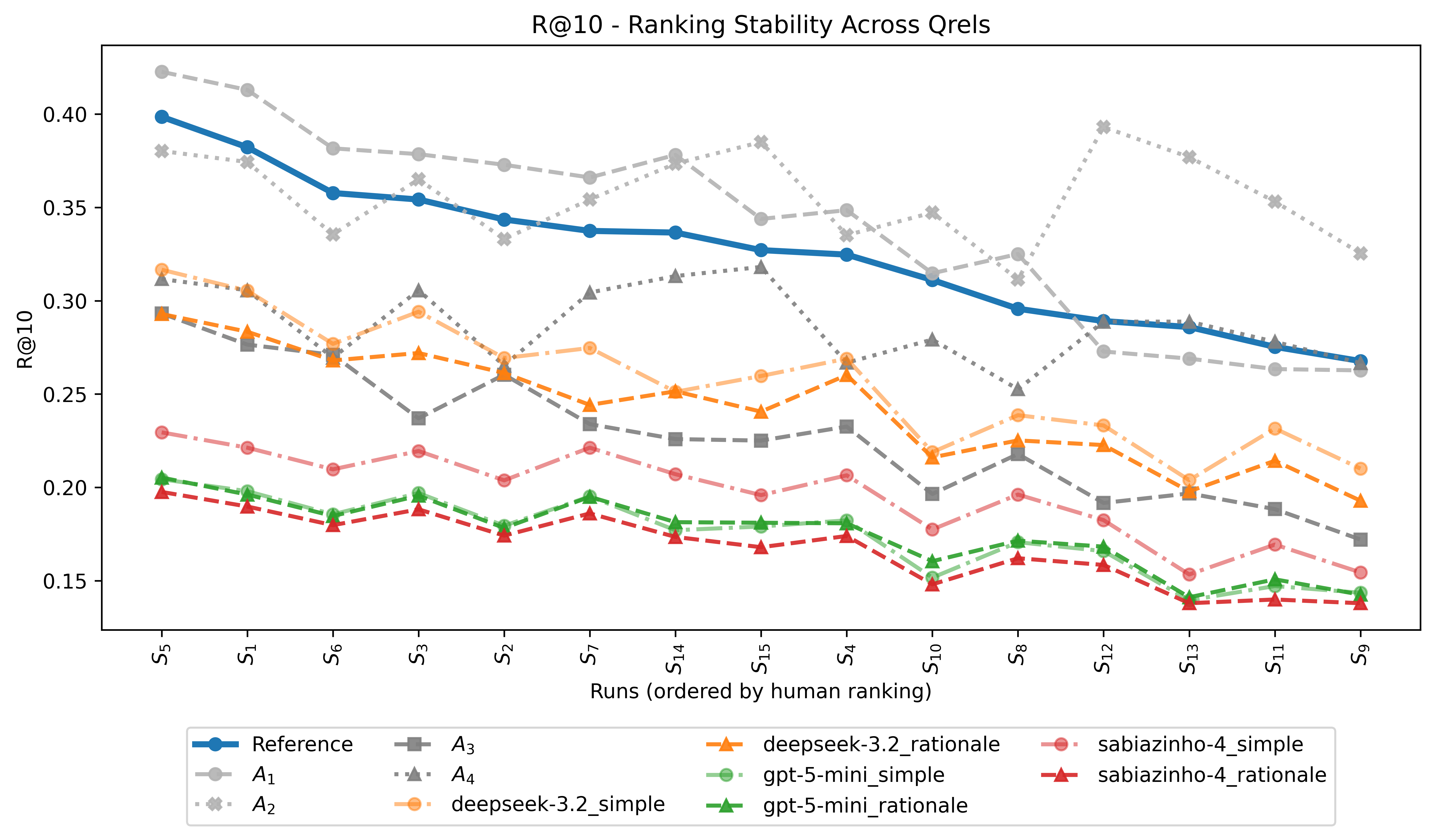}
    \caption{R@10 for each IR system computed with different qrels. Systems are ordered according to the ranking produced by the reference qrels.}
\label{fig:ranking_across_qrels_R_10}
\end{figure}

\begin{figure}[!htbp]
    \centering
    \includegraphics[width=1\linewidth]{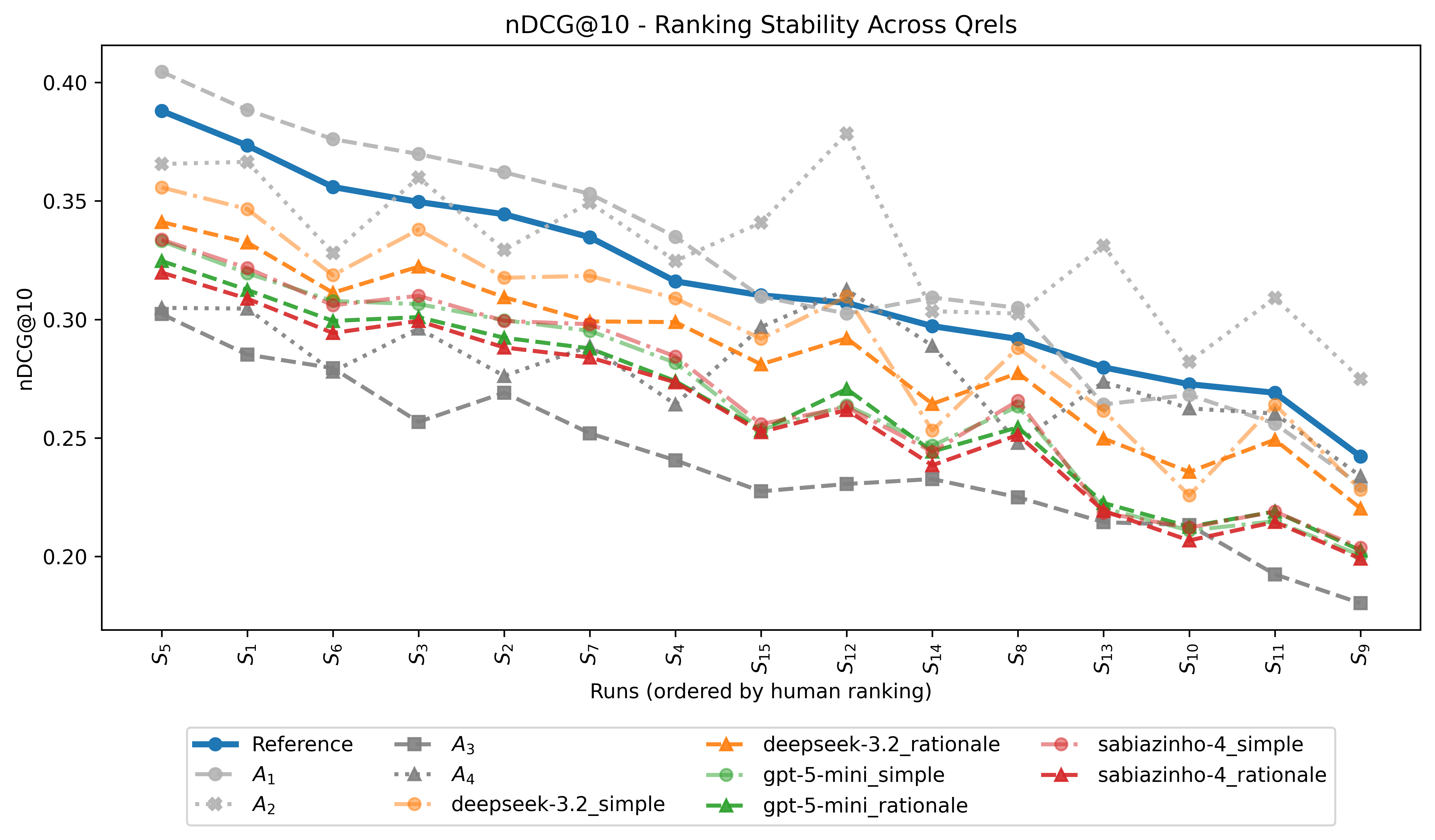}
    \caption{nDCG@10 for each IR system computed with different qrels. Systems are ordered according to the ranking produced by the reference qrels.}
\label{fig:ranking_across_qrels_nDCG_10}
\end{figure}

\begin{figure}[!htbp]
    \centering
    \includegraphics[width=1\linewidth]{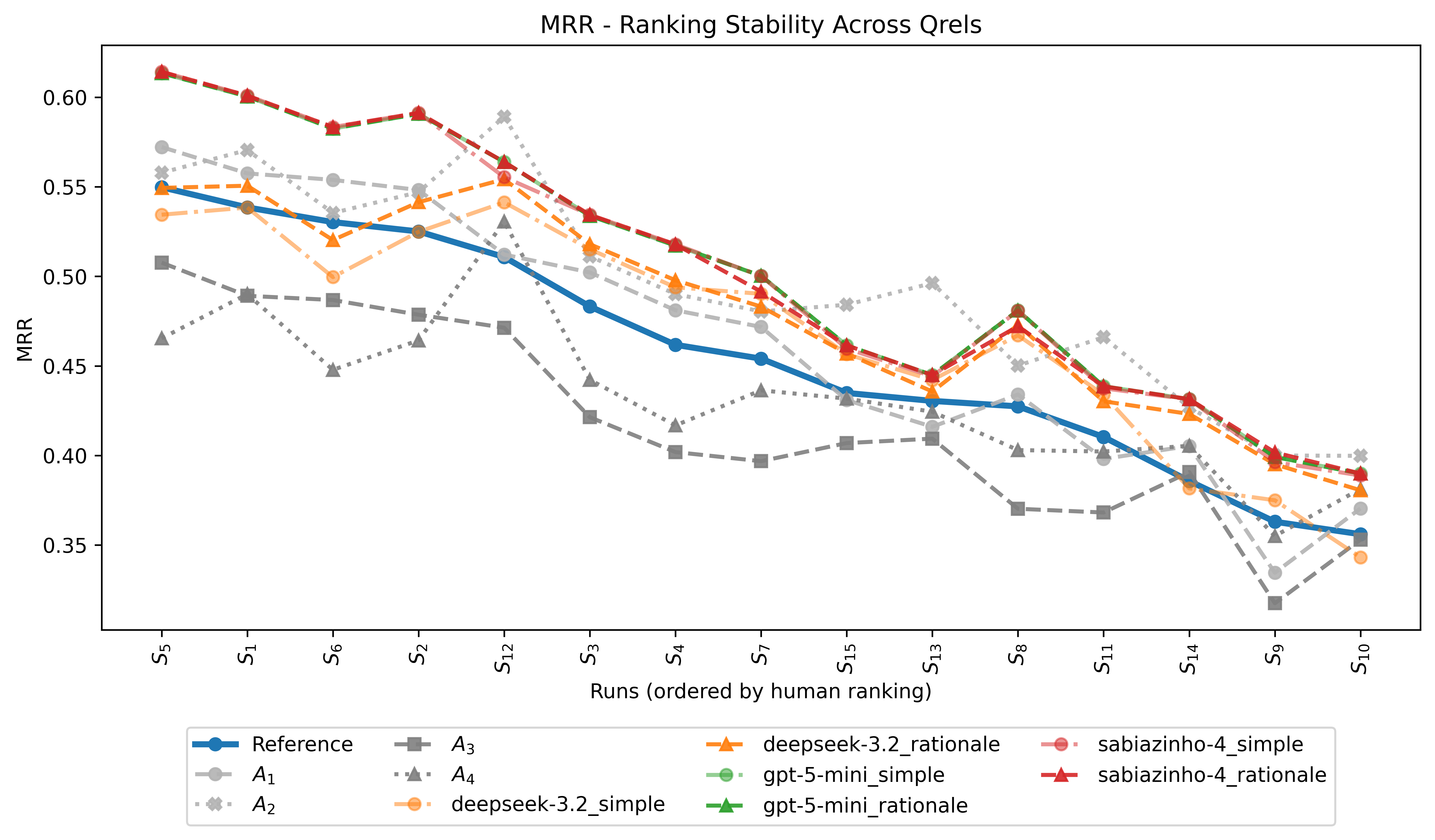}
    \caption{MRR for each IR system computed with different qrels. Systems are ordered according to the ranking produced by the reference qrels.}
\label{fig:ranking_across_qrels_MRR}
\end{figure}

Figures~\ref{fig:ranking_across_qrels_P_10}, \ref{fig:ranking_across_qrels_R_10}, \ref{fig:ranking_across_qrels_nDCG_10}, and \ref{fig:ranking_across_qrels_MRR} show P@10, R@10, nDCG@10, and MRR for each IR system evaluated with different qrels. The IR systems are ordered on the x-axis according to the corresponding metric computed using the reference qrels. Under perfect correlation, all lines would decrease monotonically. Although the lines generally decrease, they do not do so monotonically, indicating that some IR systems change their relative ranking depending on the qrels used and the metrics evaluated.

\FloatBarrier

\end{document}